\documentclass[a4paper,12pt,latin, onecolumn]{article}
\pdfoutput=1
\usepackage{amsmath}
\usepackage{subfigure}
\usepackage{graphicx}
\usepackage[english]{babel}
\usepackage[babel]{csquotes}
\usepackage{indentfirst}
\usepackage{amsmath,amssymb}
\usepackage[version=3]{mhchem}
\usepackage{booktabs}
\usepackage{color}
\usepackage{upgreek,xspace}
\usepackage[justification=justified, labelfont=bf]{caption}
\usepackage{setspace}
\usepackage{fancyhdr}
\usepackage{microtype}
\usepackage{gensymb}
\usepackage{authblk}
\usepackage{etoolbox}
\usepackage{braket}
\usepackage{lmodern}
\usepackage [a4paper, top=2.5cm, bottom=2.5cm, left=2cm, right=2cm, bindingoffset=0mm] {geometry}
\newcounter{firstbib}

\begin{document}
\doublespacing

\makeatletter
\patchcmd{\@maketitle}{\LARGE \@title}{\fontsize{18}{19.2}\selectfont\@title}{}{}
\makeatother

\renewcommand\Authfont{\fontsize{12}{13.4}\selectfont}
\renewcommand\Affilfont{\fontsize{10}{10.8}\itshape}

\title{Strongly bound excitons in anatase TiO$_2$ single crystals and nanoparticles}

\author[1]{\mbox{Edoardo Baldini}}
\author[2,3]{\mbox{Letizia Chiodo}}
\author[4]{\mbox{Adriel Dominguez}}
\author[5]{\mbox{Maurizia Palummo}}
\author[6]{\mbox{Simon Moser}}
\author[7]{\mbox{Meghdad Yazdi-Rizi}}
\author[1]{\mbox{Gerald Aub\"{o}ck}}
\author[7]{\mbox{Benjamin P. P. Mallett}}
\author[8]{\mbox{Helmuth Berger}}
\author[8]{\mbox{Arnaud Magrez}}
\author[7]{\mbox{Christian Bernhard}}
\author[6]{\mbox{Marco Grioni}}
\author[4, 9]{\mbox{Angel Rubio}}
\author[1]{\mbox{Majed Chergui}\vspace{1cm}}

\affil[1]{Laboratory of Ultrafast Spectroscopy, ISIC and Lausanne Centre for Ultrafast Science (LACUS), \'Ecole Polytechnique F\'ed\'erale de Lausanne (EPFL), CH-1015 Lausanne, Switzerland}
\affil[2]{Unit of Nonlinear Physics and Mathematical Modeling, Department of Engineering, Universit\`a Campus Bio-Medico di Roma, Via Alvaro del Portillo 21, I-00128, Rome, Italy}
\affil[3]{Center for Life Nano Science, Sapienza, IIT, Viale Regina Elena 291, I-00161, Rome, Italy}
\affil[4]{Max Planck Institute for the Structure and Dynamics of Matter, Hamburg, Germany}
\affil[5]{Dipartimento di Fisica and INFN, Universit\`a ``Tor Vergata", Via della Ricerca Scientifica 1, 00133 Roma, Italy}
\affil[6]{Laboratory of Electron Spectroscopy, IPHYS and and Lausanne Centre for Ultrafast Science (LACUS), \'Ecole Polytechnique F\'ed\'erale de Lausanne (EPFL), CH-1015 Lausanne, Switzerland}
\affil[7]{Department of Physics, University of Fribourg, Chemin du Mus\'ee 3, CH-1700 Fribourg, Switzerland}
\affil[8]{Crystal Growth Facility, \'Ecole Polytechnique F\'ed\'erale de Lausanne (EPFL), CH-1015 Lausanne, Switzerland}
\affil[9]{Departamento Fisica de Materiales, Universidad del Pa\'is Vasco, Av. Tolosa 72, E-20018, San Sebastian, Spain}

\date{\vspace{-5ex}}
\maketitle 
\newpage

\label{Introduction}

\textbf{Anatase TiO$_2$ is among the most studied materials for light-energy conversion applications, but the nature of its fundamental charge excitations is still unknown. Yet it is crucial to establish whether light absorption creates uncorrelated electron-hole pairs or bound excitons and, in the latter case, to determine their character. Here, by combining steady-state angle-resolved photoemission spectroscopy and spectroscopic ellipsometry with state-of-the-art \textit{ab initio} calculations, we demonstrate that the direct optical gap of single crystals is dominated by a strongly bound exciton rising over the continuum of indirect interband transitions. This exciton possesses an intermediate character between the Wannier-Mott and Frenkel regimes and displays a peculiar two-dimensional wavefunction in the three-dimensional lattice. The nature of the higher-energy excitations is also identified. The universal validity of our results is confirmed up to room temperature by observing the same elementary excitations in defect-rich samples (doped single crystals and nanoparticles) via ultrafast two-dimensional deep-ultraviolet spectroscopy.}

\newpage

The field of excitonics has gained increased attention in the last years, due to the unique properties that excitons manifest in the conversion and transport of energy. Key to these developments is the ability to exploit exciton physics in materials that are easily fabricated and widely available. Anatase TiO$_2$ belongs to a class of solids with superior functionalities for the conversion of light into other forms of energy \cite{ref:fujishima, ref:oregan, ref:pelizzetti}, but despite the wide effort dedicated to improve its optoelectronic performances, the microscopic nature of the fundamental electronic and optical excitations is still not understood. It is therefore pivotal to clarify the single- and two-particle excitation spectra of pure anatase TiO$_2$ and to establish the nature of the charge excitations produced upon photon absorption.

Two key aspects of anatase TiO$_2$ are: i) it crystallizes in a tetragonal unit cell, built on a network of corner- or edge- sharing TiO$_6$ octahedra (Fig. 1a), with a substantial difference between the lattice constants a = 3.78 \AA~and c = 9.51 \AA; ii) the Ti-$3d$ O-$2p$ orbital interactions run mainly in TiO$_2$ bilayers perpendicular to the [001] direction, and display only a minor contribution along the c-axis \cite{ref:chen_dawson}. This leads to an electronic structure with almost flat bands along the $\Gamma$-Z direction of the three-dimensional (3D) Brillouin zone (BZ) (Fig. 1b) and to a strong optical anisotropy for light polarized in the (001) plane and perpendicular to it.

First significant steps towards understanding the electronic excited states of this material were achieved by experimental probes such as angle-resolved photoemission spectroscopy (ARPES) \cite{ref:emori, ref:moser} and optical spectroscopy \cite{ref:tang_urbach, ref:hosaka, ref:sekiya}. Recent ARPES studies revealed that this material has an indirect bandgap, since the valence band (VB) maximum lies close to the X point and the conduction band (CB) minimum is at the $\Gamma$ point of the BZ \cite{ref:emori, ref:moser}; consequently, the lowest optical absorption edge can be described in terms of an Urbach tail caused by the phonon-induced localization of excitons \cite{ref:tang_urbach}. Less experimental attention, however, has been paid to the detailed characterization of the optical response above the absorption threshold, where anisotropy effects become more pronounced \cite{ref:hosaka, ref:sekiya}. In particular, the role played by many-body correlations in the optical properties has remained elusive to experimental probes, leading to a lack of knowledge about the nature of the elementary direct charge excitations in this material.

Many-body correlations have been investigated within the theoretical framework of Density Functional Theory (DFT) with perturbation-theory corrections at the G$_0$W$_0$ level. This \textit{ab initio} method provided a preliminary description of the material dielectric function \cite{ref:chiodo, ref:lawler, ref:kang, ref:landmann}, despite neglecting the roles of doping, electron-phonon coupling, temperature effects and indirect transitions. The diagonalization of the Bethe-Salpeter Hamiltonian predicted several direct optical transitions at energies well below the direct electronic gap computed at the GW level. The existence of these bound localized excitons in anatase TiO$_2$ is, however, still awaiting experimental verification, due to the difficulty of measuring the exciton binding energy ($E_\mathrm{B}$) for an indirect gap material. Indeed, conventional experimental techniques like optical absorption \cite{ref:muth}, photoluminescence \cite{ref:sell, ref:rinaldi} and magneto-optics \cite{ref:maan} are not suitable, as the onset of the direct band-to-band transitions cannot be identified; moreover, to derive $E_\mathrm{B}$, these methods often rely on approximate models valid only for Wannier-Mott excitons \cite{ref:elliott}.

The present work uses a unique combination of steady-state and ultrafast experimental tools with advanced theoretical calculations, to unambiguously reveal the role of many-body correlations in anatase TiO$_2$ and identify the nature of its elementary electronic excitations. By ARPES and spectroscopic ellipsometry (SE), we provide accurate values of the direct gap for both charged and neutral excitations. This leads us to unravel the existence of strongly bound excitons in this material and to offer a rigorous estimate of $E_\mathrm{B}$ for a direct exciton rising over the continuum of indirect (phonon-mediated) interband transitions, free from assumptions on the nature of the excitonic species under study. These results are supported by many-body perturbation-theory calculations, which include for the first time the role of doping, electron-phonon coupling and indirect transitions in this material. Our calculations confirm the stability of bound excitons and provide a complete description of their real-space behaviour. The room temperature (RT) robustness and generality of these elementary excitations is finally demonstrated by an ultrafast deep-ultraviolet (UV) two-dimensional (2D) spectroscopic study of widely different samples, ranging from single crystals with various degrees of doping to colloidal nanoparticles (NPs).

\section*{Results}
\subsection*{Angle-resolved photoemission spectroscopy}

To reveal the possible existence of a direct exciton, an accurate determination of the direct electronic bandgap is needed. To this aim, we perform ARPES measurements on anatase TiO$_2$ single crystals at 20 K, using a photon energy $h\nu$ = 128 eV and a resolution of 30 meV. In particular, we introduce an excess electron density ($n$) in the CB of the material by inducing oxygen vacancies in a (001)-oriented single crystal, as described in Ref. 6. As a consequence, the Fermi level ($E_\mathrm{F}$) pins slightly above the CB edge, providing a robust reference to measure the quasiparticle gap at the $\Gamma$ point. To evaluate the shift of $E_\mathrm{F}$ above the CB edge, in Fig. 2 we show the energy distribution curves at the X (blue line) and $\Gamma$ (red line) points of the 3D BZ for a crystal doped with an excess electron density $n$ = 2 $\times$ 10$^{19}$ cm$^{-3}$. In both spectra, the structure arising at -1 eV corresponds to the well-established in-gap oxygen defect states \cite{ref:thomas}. The curve at $\Gamma$ exhibits spectral weight at $E_\mathrm{F}$, which lies 80 meV above the conduction quasiparticle band.

The valence states at 20 K along the $\Gamma$-X line are displayed in Fig. 3a and referenced to the bottom of the CB. To resolve the dispersive features present in the ARPES maps and evaluate the energy of the quasiparticle in the VB, we use the approach based on computing the second derivative of the ARPES data with respect to the energy (Fig. 3b) \cite{ref:pzhang}. In this panel, the dashed blue line, highlighting the top of the VB, has been drawn as a guide to the eye. As expected, the VB onset occurs close to the X point, and precedes the rise at $\Gamma$ by $\sim$0.5 eV. Direct inspection of the band structure yields a first highly dispersing band close to the VB upper edges, whose maxima in the vicinity of the X and $\Gamma$ points are at -3.47 $\pm$ 0.03 eV and -3.97 $\pm$ 0.03 eV, respectively, representing valuable estimates of the quasiparticle energies.

We also focus on the $\Gamma$ point and monitor the evolution of the quasiparticle gap as a function of doping, by performing ARPES measurements at variable excess electron density (5 $\times$ 10$^{17}$ cm$^{-3}$ $\leq$ $n$ $\leq$ 5 $\times$ 10$^{20}$ cm$^{-3}$) in the CB. The energy-momentum dispersion relations for the bottom of the CB in the two extreme $n$-doping levels considered in our experiment are shown in Supplementary Figs. 3a-b; the second derivatives of these maps with respect to the energy axis are shown in Supplementary Figs. 3c-d. The latter allow us to identify the position of the quasiparticle energy at the bottom of the CB. Supplementary Fig. 3e compares the energy distribution curves at the $\Gamma$ point of the BZ at the two considered doping levels. Relying on the second derivative analysis also for the VB, we can estimate the energy difference from the quasiparticle VB maximum to the CB minimum at $\Gamma$ for the two doping levels at $\sim$3.98 eV. This shows that there is no measurable bandgap renormalization (BGR) at the $\Gamma$ point upon increased electron concentration over three orders of magnitude. This observation is also supported by many-body perturbation theory results (see below).

\subsection*{Spectroscopic ellipsometry}

As far as the direct gap of the two-particle excitation spectrum (\textit{i.e.} the optical spectrum) is concerned, a very reliable experimental technique for measuring the dielectric function $\epsilon$($\omega$) = $\epsilon_1$($\omega$) + i$\epsilon_2$($\omega$) of a material is SE. Figures 4a-b show the imaginary part of the dielectric function, $\epsilon_2$($\omega$), measured at 20 K on (010)-oriented single crystals, with light polarized perpendicular (\textbf{E} $\perp$ c) and parallel (\textbf{E} $\parallel$ c) to the c-axis, respectively. The spectra are obtained both on a pristine ($n$ $\sim$ 0 cm$^{-3}$) crystal (blue lines) and on the same $n$-doped crystal used for the ARPES measurement ($n$ = 2 $\times$ 10$^{19}$ cm$^{-3}$) (red lines). In the pristine crystal, the direct absorption for \textbf{E} $\perp$ c (Fig. 4a) is characterized by the presence of a sharp peak at 3.79 eV (I), preceded by a long Urbach tail at lower energies \cite{ref:tang_urbach}. A second, broader charge excitation (II) lies at 4.61 eV and extends up to 5.00 eV. The c-axis response (Fig. 4b) is instead characterized by a feature peaking at 4.13 eV (III) with a shoulder at 5.00 eV. Remarkably, all these excitations are still clear-cut in the $n$-doped sample, where we observe: i) no apparent shift in the peak energy of features I and III and an 80 meV redshift in the peak energy of feature II; ii) a reduction of the oscillator strength of all peaks, due to a transfer of spectral weight from the above-gap to the below-gap region; iii) a pronounced broadening of the spectral features.

From Fig. 4, we conclude that the lowest direct optical excitation, \textit{i.e.} the direct optical gap of the material, is the feature at 3.79 eV for both pristine and $n$-doped ($n$ = 2 $\times$ 10$^{19}$ cm$^{-3}$) anatase TiO$_2$. Combined with the ARPES data, we also conclude that it belongs to a bound exciton of the $n$-doped single crystal, with an $E_\mathrm{B}$ = 180 meV, which is experimentally derived for the first time. We also investigate the temperature effect on the stability and renormalization of exciton I, and compare the dielectric function of pristine anatase TiO$_2$ at 20 K and 300 K (Supplementary Fig. 6). Peak I is found to undergo a blueshift of 40 meV from 20 K to 300 K, while peak II undergoes a small redshift. The detailed interpretation of the full temperature dependence will be the subject of a separate publication, but here we briefly stress that this observation implies the stability of exciton I even at RT. In addition, the blueshift of this peak is anomalous \cite{ref:cardona}, because it is opposite to the trends observed in standard insulators \cite{ref:varshni}. Blueshifts of high-energy electronic excitations have previously been reported in only few materials, and explained phenomenologically by invoking different aspects related to the electron-phonon coupling: The temperature dependence of the Debye-Waller factors in PbTe \cite{ref:keffer}, the \textit{p}-\textit{d} hybridization modulated by the electron-phonon interaction in chalcopyrites \cite{ref:pwyu, ref:bhosale}, the Fr\"ohlich interaction in perovskite titanates \cite{ref:rossle}. We will show below that, by taking into account the role of the electron-phonon coupling and temperature effects, our \textit{ab initio} calculations are able to reproduce the anomalous blueshift of exciton peak I.

\subsection*{Many-body perturbation theory calculations}

To rationalize our experimental results, we perform extensive \textit{ab initio} calculations, including for the first time the effects of a finite doping and of the electron-phonon coupling on both the electronic and optical response of anatase TiO$_2$. Combined with highly converged results for the pristine crystal, such calculations allow us to assess the role of doping and temperature via a more realistic model of the material and to establish a direct comparison with the ARPES and SE data. First, we calculate the pristine anatase TiO$_2$ GW band structure within the frozen lattice approximation. Figure 5 shows the complete band structure of pristine anatase TiO$_2$. Grey diamonds denote the values obtained within GW for the VB and CB at the $\Gamma$ and X points. These are also displayed in Fig. 3b for a direct comparison with the ARPES data. The overall agreement is good, but the theoretical gap values are higher by $\sim$100 meV. This discrepancy can be caused by the effects of the doped electron density in the experimental data and/or the presence of strong electron-phonon interaction in anatase TiO$_2$. In the following, we address the relevance of these separate effects.

In general, in the presence of doping, two competing effects contribute to changing the electronic gap of an insulator, resulting in an overall redshift or blueshift depending on which effect dominates. The CB filling gives rise to a blueshift, while BGR (with the system becoming slightly metallic) is responsible for a redshift as a result of exchange and correlation effects. To address the impact of the doped carrier density on the electronic band structure, the GW theoretical framework has been extended to the case of uniformly doped anatase TiO$_2$. We observe that the change in the electronic direct gap at the $\Gamma$ point is marginally affected by doping a uniform excess electron density for $n$ $\leq$ 10$^{20}$ cm$^{-3}$. The calculated GW gaps (both  direct and indirect gap) are similar to the pristine anatase TiO$_2$ case, displaying only a slight increase well below the computational resolution (20 meV). This also implies that the dominant effect in anatase TiO$_2$ for the considered doping range is the CB filling. These results are consistent with the experimental ARPES data of Supplementary Fig. 3 and show that the electronic gap of doped samples delivers a suitable value of the gap of pristine anatase TiO$_2$. For this reason, hereafter we refer only to the results of the GW calculations in the pristine sample. We will later show that the same holds for the optical response, as the position and shape of peak I in the calculated dielectric function do not change for $n$ = 10$^{19}$ cm$^{-3}$. We therefore rule out the doped electron density as origin of the discrepancy between the experimental ARPES data and the calculations.

Thus, it is crucial to estimate the role of the electron-phonon interaction, which is known to be relatively strong in anatase TiO$_2$ \cite{ref:moser, ref:toyozawa, ref:deskins, ref:divalentin, ref:jacimovic, ref:setvin, ref:tang_PL}. To this aim, we consider both zero point renormalization (ZPR) and electron-phonon coupling at finite temperature. For the former, we rely on recent theoretical data for rutile TiO$_2$ \cite{ref:monserrat}, where a redshift of 150 meV was estimated for the electronic gap. Assuming a similar correction for anatase TiO$_2$, we estimate a theoretical gap of 3.92 eV at zero temperature. To account for the electron-phonon coupling at finite temperature, frozen-phonon GW calculations for 20 K and 300 K are performed. Specifically, we calculate the electronic bandgap when the ions in the primitive unit cell are displaced according to the eigenvectors of the infrared-active $\mathrm{E_{u}}$ and $\mathrm{A_{2u}}$ normal modes \cite{ref:gonzalez}, which are the ones most strongly coupled to the electronic degrees of freedom \cite{ref:moser, ref:deskins}. We additionally consider the effect of lattice thermal expansion, which is also a source of renormalization for the quasiparticle gap. We find that the combined effect of the lattice expansion and electron-phonon coupling leads to a net blueshift of 30 to 50 meV at 300 K, while the shift is negligible at 20 K. Hence, at low temperature, the change in the electronic bandgap of anatase TiO$_2$ is only due to the ZPR. This leads to a theoretical value of the direct bandgap of 3.92 eV, which is in excellent agreement with the experimental value of 3.97 $\pm$ 0.03 eV.

The complete GW band structure in Fig. 5 further yields important information about the $\Gamma$-Z direction, which is crucial for understanding the optical transitions of the material. The CB and VB dispersions between $\Gamma$ and Z are nearly parallel, providing a large joint density of states for the optical transitions (shown as violet arrows). As discussed below, this peculiar dispersion leads to the intense excitonic transitions observed in the optical absorption spectrum.

To identify the microscopic nature of the optical excitations, we calculate $\epsilon_2$($\omega$) for both pristine and doped anatase TiO$_2$ at zero temperature. The effects of electron-phonon interactions are also included via frozen-phonon calculations at 20 K and 300 K. The results for the pristine crystal at zero temperature, with and without many-body electron-hole correlations, are shown in Fig. 6a-b. The optical spectra in the uncorrelated-particle picture (red lines) are obtained within the random-phase approximation (RPA) on top of GW, while the many-body optical spectra (violet lines) are calculated by solving the Bethe-Salpeter Equation (BSE) as implemented in the BerkeleyGW code \cite{ref:deslippe} (see Methods and Supplementary Note 4). The inclusion of many-body effects \cite{ref:chiodo, ref:lawler, ref:kang, ref:landmann} yields an excellent agreement with the SE spectra. For \textbf{E} $\perp$ c, the sharp absorption maximum at 3.76 eV is very close to feature I (3.79 eV) and well below the direct VB-to-CB optical transition evaluated at the independent particle level (3.92 eV, red trace). A second peak at 4.81 eV clearly corresponds to the experimental peak II (4.61 eV). For \textbf{E} $\parallel$ c, the intense optical peak at 4.28 eV is easily assigned to the experimental peak III (4.13 eV). The BSE calculations performed for doped anatase TiO$_2$ at concentrations of $n$ = 10$^{19}$ cm$^{-3}$ and $n$ = 10$^{20}$ cm$^{-3}$ (see Supplementary Note 6 and Supplementary Fig. 13) show a negligible effect on the considered optical peaks, supporting our SE data (Fig. 4a). The insensitivity of the energy of peak I with doping concentration suggests that effects of long-range Coulomb screening and BGR perfectly compensate each other for this transition, even at high doping levels.

\noindent We also investigate the role of lattice thermal expansion and electron-phonon interaction in the absorption spectra, by carrying out frozen-phonon BSE simulations, described in Supplementary Note 7. The position of the excitonic peak I at 20 K shows a negligible shift with respect to the zero temperature value and blueshifts by 75 meV due the temperature increase from 20 K to 300 K, in agreement with the experiment (Supplementary Fig. 6).

\noindent Finally, although exciton I is bound with respect to the direct electronic gap, it may retain a resonant character with respect to all phonon-mediated indirect transitions of the material. Hence, to account for possible effects originating from the indirect nature of the anatase TiO$_2$ gap, we solve the BSE for a supercell, thus allowing for the coupling of electron and hole states via phonons with non-zero momenta. We find a negligible role of the phonon-mediated transitions in the exciton properties of anatase TiO$_2$, beyond adding an Urbach tail at the lower energy side of peak I (see Supplementary Note 8 for details).

\subsection*{Exciton isosurface and binding energy}
The analysis in real and reciprocal space reveals very insightful information about the microscopic nature of the excitons. As far as the exciton associated with peak I is concerned, it extends two-dimensionally in a single (001) atomic plane. By fitting the exciton wavefunction with a 2D hydrogen model, we estimate the exciton Bohr radius around 3.2 nm, the 90$\%$ of the excitonic squared modulus wavefunction being contained within 1.5 nm~ (Fig. 7a and 7b). In addition, a reciprocal space analysis shows that this wavefunction is formed mainly by mixing of single-particle vertical transitions along the $\Gamma$-Z direction (Fig. 5). Due to the almost parallel CB and VB dispersion along $\Gamma$-Z, the electronic gap at $\Gamma$ (coincident with the continuum absorption rise) is used as a reference to evaluate $E_\mathrm{B}$. Considering the renormalization of the electronic bandgap discussed above, we estimate a theoretical value of $E_\mathrm{B}$ = 160 meV at 20 K, in line with our measurements.

The energy, shape and reciprocal space contributions of peak II highlight its bulk-resonance character, most evident as its offset coincides with the independent particle-GW absorption rise. Figure 7c depicts the spatial distribution of the charge excitations associated with peak II, showing significant contributions from extended bulk states.  Its electron wavefunction appears completely delocalized over many lattice constants
around the hole. Hence, this peak is assigned to a resonant excitation that does not form a bound state. Carrying out a similar analysis for peak III, we conclude that it presents a mix of localized and bulk-resonant contributions, as the continuum onset in independent particle-GW undergoes an intensity enhancement. Different from peak I, the $k$-points contributing to this charge excitation are both located along the $\Gamma$-Z line and distributed in the whole BZ. Indeed, as shown in Fig. 7d, the linear combination of excitonic wavefunctions (with eigenvalues in the range 4.18-4.38 eV for \textbf{E} $\parallel$ c) contributing to peak III leads to a fairly localized exciton in all three directions. By fitting only the bound contribution to the wavefunction with a 3D hydrogen model, we obtain an average radius of 0.7 nm, corresponding to less than two lattice constants in the plane. By including also the contributions from resonant states, the 90$\%$ of the exciton wavefunction square modulus is contained within 2 nm. Due to the mixed nature of exciton III, it is less straightforward to define and estimate its $E_\mathrm{B}$. Assuming the independent particle-GW onset at 4.40 eV for \textbf{E} $\parallel$ c as the reference energy (marked as $E_\mathrm{{dir}}$ in Fig. 6b), we obtain $E_\mathrm{B}$ $\sim$150 meV (in the frozen lattice scheme), which is still well above the standard $E_\mathrm{B}$ observed in bulk semiconductors. From this analysis, we conclude that excitons I and III, due to their $E_\mathrm{B}$ and spatial nature, retain an intermediate character between Frenkel and Wannier-Mott regimes. On the other hand, peak II corresponds to a resonance that does not form a bound exciton.

\subsection*{Exciton physics in anatase TiO$_2$ nanoparticles}

The above description dealt with bulk single crystals of anatase TiO$_2$, but in most applications \cite{ref:fujishima, ref:oregan, ref:pelizzetti}, defect-rich samples are used at RT and ambient pressure (\textit{e.g.} NPs or mesoporous films), and one may therefore question the validity of the above conclusions to the actual systems used in applications. Indeed, one could expect that the carriers released at defects and the local electric fields generated by charged impurities would screen the Coulomb interaction in the exciton, leading to the cancellation of the binding forces. Moreover, strong exciton-defect and exciton-impurity scattering can also cause an extreme broadening of the exciton linewidth, hiding the characteristic exciton feature into the continuum of indirect interband excitations. These ideas are reinforced by the equilibrium absorption spectrum of colloidal anatase TiO$_2$ NPs (of unknown doping), which does not show obvious signatures of excitonic transitions \cite{ref:serpone, ref:monticone} (Supplementary Fig. 1 and black trace in Fig. 8a). However, this spectrum is also strongly affected by the scattering of the incident light and this can in turn affect the detection of the excitonic peaks. To circumvent this problem, we interrogate the system out-of-equilibrium via ultrafast 2D UV transient absorption spectroscopy \cite{ref:aubock}, since this technique subtracts the scattered light and provides a better contrast for resolving hidden features. It is applied here for the first time to solid samples and it offers the capability to excite across the gap of a wide-gap material with tunable photon energy and probe in the same region with a broadband continuum. Typically, the exciton lineshapes can be identified through the pump-induced transparency of the excitonic peak, referred to as “exciton bleaching”. This nonlinear optical process is intrinsically related to a many-body phenomenon: Its manifestation depends not only on the final-state interactions of electron and hole involved in the excitonic state but also on the interaction with all other particles in the material, that can contribute to screening or blocking the excitonic transition \cite{ref:haug, ref:schmitt-rink}. 

We carry out 2D UV spectroscopy of doped single crystals ($n$ $\sim$ 0 cm$^{-3}$, $n$ = 2 $\times$ 10$^{17}$ cm$^{-3}$ and $n$ = 2 $\times$ 10$^{19}$ cm$^{-3}$) in transient reflectivity ($\Delta R/R$) and of a colloidal solution of NPs in transient absorption ($\Delta A$). For the single crystals, we retrieve the anisotropic $\Delta R/R$ dynamics along the a- (Supplementary Figs. 8a-b) and c-axis (Supplementary Figs. 8c-d and 9) in a broad UV range, using a pump photon energy of 4.40 eV. Subsequently, we extract $\Delta A$ from the $\Delta R/R$ response (see Supplementary Note 3). Figures 8a and 8b respectively show the normalized $\Delta A$ of anatase TiO$_2$ NPs in aqueous solution and of doped anatase TiO$_2$ single crystals ($n$ = 2 $\times$ 10$^{19}$ cm$^{-3}$), as a function of the probe photon energy, at a time delay of 1 ps and for different pump photon energies. Figure 8a exhibits a negative (bleach) signal over the entire probe range, displaying two prominent features at 3.88 eV and 4.35 eV. While the former excitation is present at all pump photon energies, the latter becomes more prominent for pump photon energies beyond 4.10 eV, \textit{i.e.} in correspondence with the threshold for accessing the c-axis optical response. Both features have a full-width at half maximum of 300 $\pm$ 20 meV. Figure 8b shows the derived $\Delta A$ along the a- and c-axis for the doped single crystal, under 4.40 eV excitation and at a time delay of 1 ps. In these separate polarization channels, two negative features appear at 3.88 eV and 4.32 eV, similarly to the $\Delta A$ spectra of NPs. These energies match those of the excitonic peaks I and III in the equilibrium absorption spectrum along the a- and c-axis (Supplementary Figs. 4c-d, note that the values of the steady-state absorption peaks slightly differ from the peaks in the dielectric function). Hence, this analysis leads us to disentangle the a-axis and c-axis contributions in the $\Delta A$ response of the NPs. Both excitons are found to appear in the latter due to the random orientation of NPs in solution. Therefore, we conclude that the excitonic features are also present in the equilibrium absorption spectrum of anatase TiO$_2$ NPs (Supplementary Fig. 1), but they are obscured by the strong light scattering. This highlights the ability of ultrafast 2D deep-UV spectroscopy to reveal hidden features in the spectral response of wide-gap materials.

\section*{Discussion}
Our extensive study allows us to demonstrate the stability of bound excitons in anatase TiO$_2$. Excitonic effects have been thought to be weak in this material, due to its large static dielectric constant ($\epsilon_\mathrm{{S}}$ $\sim$ 22/45) \cite{ref:gonzalez, ref:roberts}. Moreover, due to a smaller effective mass \cite{ref:tang_electrical}, one would expect the electron-hole interaction to be even weaker than that in rutile TiO$_2$, in which $E_\mathrm{B}$ has been estimated $\sim$4 meV \cite{ref:pascual, ref:amtout}. However, the presence of a large $\epsilon_\mathrm{{S}}$ is not a sufficient condition to prevent the formation of excitons in materials, as the screening of the electron-hole interaction should rigorously take into account the momentum- and energy-dependence of the dielectric constant \cite{ref:toyozawa}. Therefore, it is the combination of the electronic structure and the nature of the screening that determines the existence of bound excitons in materials. The degree of excitonic spatial delocalization is instead influenced by the crystal structure, as the packing of the polyhedra containing the atoms involved in the excitonic transitions is related to the details of the band structure. These observations rationalize well the exciton physics in titanates. 

\noindent A necessary condition for a many-state transition to occur and for a bound collective excitonic state to form is that the electron and hole group velocities are the same, \textit{i.e.} that the gradients of the lowest CB and the highest VB are identical in a specific portion of the BZ \cite{ref:toyozawa}. As a matter of fact, the band edge states of anatase TiO$_2$ (Fig. 5) and SrTiO$_3$ \cite{ref:benrekia, ref:sponza} are parallel in extended portions of the BZ, which contribute density of states to the collective transition associated with bound excitonic species ($E_\mathrm{B}$ $\sim$ 180/220 meV) \cite{ref:chiodo, ref:gogoi}. In contrast, the band structure of rutile TiO$_2$ is not characterized by such extended portions with similar electron and hole group velocities \cite{ref:chiodo}, thus resulting in $E_\mathrm{B}$ $\sim$4 meV \cite{ref:pascual, ref:amtout}.

\noindent Concerning the spatial distribution of the excitons in SrTiO$_3$ and TiO$_2$, since all the bound excitonic transitions are predicted to involve O $2p$ and Ti $3d$ ($t_{2g}$) states \cite{ref:chiodo, ref:kang, ref:gogoi}, one should consider the role of different TiO$_6$ octahedra packing. In SrTiO$_3$, the unit cell is built on a distorted perovskite structure, thus providing a high degree of coordination among neighbouring TiO$_6$ octahedra. As a consequence, the single-particle band states involved in the excitonic transition undergo pronounced dispersion and the strongly bound exciton predicted in SrTiO$_3$ retains a highly delocalized nature similarly to a Wannier-Mott exciton \cite{ref:gogoi}. In TiO$_2$, both the anatase and rutile polymorphs are also built on a network of coordinated TiO$_6$ octahedra, but they significantly differ in their structural properties. In rutile, each distorted TiO$_6$ octahedron is connected to ten neighbouring ones, sharing a corner or an edge. In anatase, the coordination of the TiO$_6$ octahedra is less compact and each octahedron is coordinated only with eight neighboring ones. This seemingly tiny difference has in turn profound effects on the spatial properties of the elementary charge excitations: While in rutile the weakly-bound excitons are Wannier-Mott-like, in anatase the bound exciton is confined to the (001) plane \cite{ref:chiodo}. More specifically, it is the chain-like structure of anatase TiO$_2$ that leads to unique characteristics of the band structure and hinders the delocalization in 3D of the bound exciton over many unit cells. Indeed, the band dispersion along the $\Gamma$-Z direction is rather flat, thus implying a high degree of localization for the excitonic state along the c-axis. This almost 2D wavefunction also contributes to enhance the exciton $E_\mathrm{B}$, in a way similar to the low-dimensional effect in semiconductor quantum structures \cite{ref:schmitt-rink}. Under these conditions, according to the 2D hydrogen model, $E_\mathrm{B}$ is 4 times larger than for a 3D exciton. With $E_\mathrm{B}$ exceeding the highest energy of the infrared-active longitudinal optical phonons (\textit{i.e.} 108 meV in anatase TiO$_2$) \cite{ref:gonzalez}, the ionic (polaronic) contribution to the screening of the Coulomb interaction is strongly suppressed \cite{ref:toyozawa}. As a consequence, the total screening reduces to the pure contribution of the background valence electrons, embodied by the rather small dielectric constant at optical frequencies $\epsilon_\mathrm{{opt}} \sim$ 6/8. This weak screening in turn reduces the exciton Bohr radius on the (001) plane ($\sim$3.2 nm).

\noindent Following these arguments, the 2D wavefunction of the exciton I in anatase TiO$_2$ represents a peculiar and fascinating object. Indeed, in recent years, several 2D excitons were reported in materials such as hexagonal boron nitride \cite{ref:galambosi} and transition metal dichalcogenides \cite{ref:qiu, ref:he, ref:ugeda}, which are layered 2D systems held together by van der Waals forces to form a 3D lattice. The situation of anatase TiO$_2$ is radically different in that a genuine 3D material exhibits a 2D exciton wavefunction on a specific lattice plane.

From the point of view of devices, the newly discovered excitons may provide a significant source of optical nonlinearity, paving the way to the development of electro-optical or all-optical switches in the UV. Also engineered nanostructures exposing a large percentage of (001) facets can be useful in guiding the energy at the nanoscale in a selective way \cite{ref:yang, ref:giorgi, ref:palummo}. Finally, due to the important contribution that phonons have on the exciton width and lineshape, we expect that the optical properties of anatase TiO$_2$ can be effectively altered by tuning the exciton-phonon coupling, \textit{e.g.} through the applications of mechanical strain. In this regard, new insights from many-body theory will be crucial for evaluating the transport of these excitonic species, their coupling to the vibrational degrees of freedom and their reaction to various external stimuli.

\section*{Methods}

\subsection*{Single crystal growth and characterization}
\label{Methods_SingleCrystals}

High-quality single crystals of anatase TiO$_2$ were produced by a chemical transport method from anatase powder and NH$_4$Cl as transport agent, similar to the procedure described in Ref. 55. In detail, 0.5 g of high-purity anatase powder were sealed in a 3 mm thick, 2 cm large and 20 cm long quartz ampoule together with 150 mg of NH$_4$Cl, previously dried at $60\,^{\circ}\mathrm{C}$ under dynamic vacuum for one night, and 400 mbar of electronic grade HCl. The ampoules were placed in a horizontal tubular two-zone furnace and heated very slowly to $740\,^{\circ}\mathrm{C}$ at the source, and $610\,^{\circ}\mathrm{C}$ at the deposition zone. After two weeks, millimeter-sized crystals with a bi-pyramidal shape were collected and cut into rectangular bars (typically 0.8 $\times$ 0.6 $\times$ 0.15 $\mathrm{mm^3}$). Copper-doped anatase TiO$_2$ single crystals were obtained by annealing raw anatase single crystals in O$_2$ at $700\,^{\circ}\mathrm{C}$ for 6 days in the presence of Cu vapors. The pristine form of anatase TiO$_2$ was instead obtained by annealing the raw anatase TiO$_2$ crystals at $700\,^{\circ}\mathrm{C}$ for 10 days under 950 mbar of CO. The doping levels of the raw, copper-doped and pristine crystals were determined via ARPES or transport measurements to be $n$ = 2 $\times$ 10$^{19}$ cm$^{-3}$, $n$ = 2 $\times$ 10$^{17}$ cm$^{-3}$ and $n$ $\sim$ 0 cm$^{-3}$, respectively.

\subsection*{Nanoparticles synthesis and characterization}
\label{Method_NPs}

Anatase TiO$_2$ NPs were prepared by the sol-gel method \cite{ref:mahshid}. The synthesis was carried out in a glove box under argon atmosphere. Titanium isopropoxide (Sigma Aldrich, 99.999$\%$ purity) was used as precursor and mixed with 10 ml of 2-propanol. This mixture was added dropwise under vigorous stirring to cold acidic water ($2\,^{\circ}\mathrm{C}$, 250 ml H$_2$O, 18 M$\Omega$, mixed with 80 ml glacial acetic acid, final pH 2). At the beginning the mixture looked turbid, but after stirring it in an ice bath for 12 hours, it became transparent as the amorphous NPs were formed. Half of the mixture was left stirring for days to stabilize the NPs. The other half was peptized at $80\,^{\circ}\mathrm{C}$ for about 2 hours until the liquid turned into a transparent gel. The gel was autoclaved at $230\,^{\circ}\mathrm{C}$ for 12 hours. During this process the previous amorphous sample became denser and underwent a phase transition, resulting in anatase TiO$_2$ NPs. After the autoclave, the NPs precipitated to the bottom of the container. They were separated from the supernatant and added to 100 ml acidic water (pH 2) to obtain a white colloidal solution with a final concentration of ca. 337 mM. In Refs. 57-58, we reported the details of the sample characterization by means of X-ray diffraction and transmission electron microscopy. These techniques enabled us to demonstrate the good quality of the anatase phase and the spherical shape (with an average diameter of approximately 25 nm) of the NPs. The doping of the NPs was not estimated and, thus, it is unknown. The steady-state absorption spectrum of the colloidal solution of anatase TiO$_2$ was recorded at RT using a commercial UV-VIS-NIR spectrometer (Shimadzu, UV-3600). Before measuring the absorption spectrum of the sample, a reference spectrum of the pure solvent (acidic water, pH 2) was taken to check its transparency in the investigated spectral range.

\subsection*{Angle-resolved photoemission spectroscopy}
\label{Methods_ARPES}

The ARPES measurements were performed at the Electronic Structure Factory endstation on beamline 7.0.1 at the Advanced Light Source, Berkeley, USA. A raw anatase TiO$_2$ single crystal was polished and cleaned in a buffered 5$\%$ fluoric acid solution before introducing it into the ultra-high vacuum system (\textless $10^{-10}$ mbar). The crystal was annealed in 35 mbar of oxygen at $400\,^{\circ}\mathrm{C}$ for 30 minutes before the ARPES experiments. The ARPES measurements were performed at a photon energy $h\nu$ = 128 eV and with an energy resolution of 30 meV.

\subsection*{Spectroscopic ellipsometry}
\label{Method_Ellipsometry}

Using SE, we measured the complex dielectric function of the sample, covering the spectral range from 1.50 eV to 5.50 eV. The measurements were performed using a Woollam VASE ellipsometer. The single crystals with $n$ $\sim$ 0 cm$^{-3}$ and $n$ = 2 $\times$ 10$^{19}$ cm$^{-3}$ were polished along a (010)-oriented surface and mounted in a helium flow cryostat, allowing measurements from RT down to 10 K. When at cryogenic temperatures, the measurements were performed at $<$10$^{-8}$ mbar to prevent measurable ice-condensation onto the sample. Anisotropy corrections were performed using standard numerical procedures \cite{ref:aspnes} and diffraction effects at low frequency were accounted for using the procedure developed in Ref. 60. The SE data have been further corrected to account for the surface roughness of the single crystal, which was estimated around 0.9 nm by means of Atomic Force Microscopy (AFM). Supplementary Figs. 2a-b shows two images taken under the AFM for the $n$ $\sim$ 0 cm$^{-3}$ crystal. The average surface roughness of the polished surfaces was $\sim$ 0.9 nm. The steady-state reflectance at 100 K was derived from the measured dielectric function to be compared to previous normal-incidence reflectivity measurements (Supplementary Note 1 and Supplementary Figs. 4a-b). The precision of the Kramers-Kronig analysis used to treat the normal-incidence reflectivity data was also tested (Supplementary Fig. 5)

\subsection*{Ultrafast 2D UV spectroscopy}
\label{Methods_Ultrafast}

The ultrafast optical experiments were performed using a novel set-up of tunable UV pump and broadband UV probe, described in detail in Ref. 38. A 20 kHz Ti:Sapphire regenerative amplifier (KMLabs, Halcyon + Wyvern500), providing pulses at 1.55 eV, with typically 0.6 mJ energy and around 50 fs duration, pumped a noncollinear optical parametric amplifier (NOPA) (TOPAS white - Light Conversion) to generate sub-90 fs visible pulses (1.77 - 2.30 eV range). The typical output energy per pulse was 13 $\upmu$J. Around 60\% of the output of the NOPA was used to generate the narrowband pump pulses. The visible beam, after passing through a chopper, operating at 10 kHz and phase-locked to the laser system, was focused onto a 2 mm thick BBO crystal for nonlinear frequency doubling. The pump photon energy was controlled by the rotation of the crystal around the ordinary axis and could be tuned in a spectral range up to $\sim$0.9 eV ($\sim$60 nm) wide. The typical pump bandwidth was 0.02 eV (1.5 nm) and the maximum excitation energy was about 120 nJ. The pump power was recorded on a shot-to-shot basis by a calibrated photodiode for each pump photon energy, allowing for the normalization of the data for the pump power. The remaining NOPA output was used to generate the broadband UV probe pulses with $\sim$1.3 eV ($\sim$100 nm) bandwidth through an achromatic doubling scheme. Pump and probe pulses, which have the same polarization, were focused onto the sample, where they were spatially and temporally overlapped. The typical spot size of the pump and the probe were 100 $\upmu$m and 40 $\upmu$m full-width at half-maximum respectively, resulting in a homogeneous illumination of the probed region. 

This setup could be used either in a transmission or in a reflection configuration. The anatase TiO$_2$ single crystals were studied by detecting their transient reflectivity ($\Delta R/R$) upon photoexcitation, while the NPs were investigated by recording their transient absorption ($\Delta A$). In the case of the measurements on the anatase TiO$_2$ single crystals, the specimens were mounted on a rotating sample holder, in order to explore the $\Delta R/R$ response along the desired crystalline axis. The measurements were performed on the three different classes of samples ($n$ $\sim$ 0 cm$^{-3}$, $n$ = 2 $\times$ 10$^{17}$ cm$^{-3}$ and $n$ = 2 $\times$ 10$^{19}$ cm$^{-3}$, see Supplementary Note 2 and Supplementary Fig. 7). The portion of the probe beam reflected by the surface of the crystal was detected and the time evolution of the difference in the UV probe reflection with and without the pump pulse reconstructed. All the experiments were performed at RT. Concerning the measurements on the anatase NPs, the sample consisted of anatase TiO$_2$ NPs dispersed in an aqueous solution (20\% acetic acid and 80\% water) to avoid interparticle charge-transfer. The colloidal solution circulated into a 0.2 mm thick quartz flow-cell to prevent photo-damage and its concentration was adjusted to provide an optical density of approximately 0.4. The probe was measured after its transmission through the sample and its detection synchronized with the laser repetition rate. The difference of the probe absorption with and without the pump pulse was measured at different time delays between the pump and the probe, by means of a motorized delay line in the probe path. After the sample, the transmitted/reflected broadband probe beam was focused in a multi-mode optical fiber (100 $\upmu$m), coupled to the entrance slit of a 0.25 m imaging spectrograph (Chromex 250is). The beam was dispersed by a 150 gr/mm holographic grating and imaged onto a multichannel detector consisting of a 512 pixel CMOS linear sensor (Hamamatsu S11105, 12.5 $\times$ 250 $\upmu$m pixel size) with up to 50 MHz pixel readout, so the maximum read-out rate per spectrum (almost 100 kHz) allowed us to perform shot-to-shot detection easily. The described experimental setup offered a time resolution of 150 fs.

\subsection*{\textit{Ab initio} calculations - Pristine anatase TiO$_2$}
\label{AbInitio_Details}

Many-body perturbation theory at the level of the GW and the BSE \cite{ref:hedin1, ref:hedin2, ref:onida} was employed to compute the band structure and the dielectric response of bulk anatase TiO$_2$. The GW and BSE calculations were performed on-top of eigenvalues and eigenfunctions obtained from DFT. We used the planewave pseudopotential implementation of DFT as provided by the package Quantum Espresso. GW and BSE calculations were performed with the BerkeleyGW package \cite{ref:deslippe}. We also used the GW + BSE Yambo \cite{ref:yambo} implementation to verify that the results of our calculations are code independent.

The DFT calculations were performed using the generalized gradient approximation (GGA) as in the Perdew-Burke-Ernzerhof (PBE) scheme for the exchange-correlation functional. The Ti norm-conserving pseudopotential was generated in the Rappe-Rabe-Kaxiras-Joannopoulos (RRKJ) scheme \cite{ref:RRKJ}, including semicore 3\textit{s} and 3\textit{p} states. While standard structural and electronic quantities are already converged in DFT  with an energy cutoff of 90 Ry, the energy cutoff used here was raised to 160 Ry to properly include the high number of bands necessary to reach convergence for the many-body evaluated properties. Bulk anatase TiO$_2$ was modeled on a body-centered tetragonal lattice containing 2 Ti atoms and 4 O atoms (primitive cell) with lattice parameters (optimized at the PBE level) a = b = 3.79 \AA~ and c = 9.66 \AA. The experimental lattice constants at RT are a = b = 3.78 \AA~ and c = 9.51 \AA. Scaling these parameters to zero temperature via a linear extrapolation \cite{ref:cardona_book} of the temperature dependence of the lattice constant at high temperature, appearing in Ref. 67, yields a = b = 3.78 \AA~ and c = 9.49 \AA.   

The ground state electronic density is properly described with a coarse 4$\times$4$\times$4 $k$-point grid for sampling of the BZ. The GW quasiparticle corrections to the DFT eigenvalues were performed at the one-shot level of theory (G$_0$W$_0$). For the computation of the polarizability and inverse dielectric matrices in BerkeleyGW, we employed a total of 2474 CBs and G-vectors with kinetic energies up to 46 Ry, whereas the self-energy operator was computed using 2472 unoccupied bands and a G-vector cutoff energy of 46 Ry and 160 Ry for the screened and bare Coulomb matrices, respectively. The coarse 4$\times$4$\times$4 $k$-point grid sampling is sufficient for the description of the quasiparticle corrections, while a high number of bands is mandatory to get a proper description of screening effects and many-body corrections. The electronic band structure was finally obtained by interpolating GW corrections on top of a more refined DFT calculation with a 16$\times$16$\times$16 grid. The fully converged BSE results shown in the main text were obtained with BerkeleyGW. We used a shifted grid with up to 16$\times$16$\times$16 $k$-points (4096 irreducible $k$-points). The six lowest CBs and six topmost VBs were included to solve the excitonic Hamiltonian. The results (shown in Supplementary Figs. 10a-b, 11 and 12) were code-independent, as verified by comparing the BerkeleyGW results with those obtained with the Yambo code at the same level of convergence \cite{ref:chiodo}. All results shown in this paper were obtained with the resonant part of the excitonic Hamiltonian (inclusion of the antiresonant part does not lead to significant changes). Spin-polarized calculations were performed to highlight possible dark excitons due to triplet excitations but no measurable differences with respect to the spin-restricted results were obtained. The novelty of these calculations compared to the results reported in literature is described in Supplementary Note 5.\\

\subsection*{\textit{Ab initio} calculations - Doped anatase TiO$_2$}

Within the same theoretical framework used for pristine anatase TiO$_2$, we performed calculations for the case of uniformly doped anatase TiO$_2$, to verify computationally that the influence of doping on both the electronic gap and optical response can be disregarded. In Supplementary Note 6, we report the results for two cases of uniform excess electron density $n$ = 10$^{19}$ cm$^{-3}$ and  $n$ = 10$^{20}$ cm$^{-3}$.\\

\subsection*{\textit{Ab initio} calculations - Electron-phonon coupling}

To estimate the role of the electron-phonon coupling in the electronic and optical properties of anatase TiO$_2$, we performed frozen phonon DFT + GW + BSE calculations by separately displacing the ions in the primitive unit cell according to the eigenvector of the infrared-active $\mathrm{E_{u}}$ and $\mathrm{A_{2u}}$ normal modes \cite{ref:gonzalez}, which are those possessing the stronger coupling with the electronic degrees of freedom \cite{ref:moser, ref:deskins}. The displacement of atom $j$ was calculated from the harmonic oscillator mean square displacement at 300 K according to
\begin{equation}
< |u\mathrm{_{j}}(t)|^2> =\frac{\hbar(1 + 2n_\mathrm{{BE}})}{2 m\mathrm{_j} \omega}
\end{equation}
where $n_\mathrm{{BE}} = (e^{\hbar\omega/k\mathrm{_B} T} - 1)^{-1}$ is the Bose-Einstein statistical occupation factor, T is the temperature, $k\mathrm{_B}$ is the Boltzmann constant, $m\mathrm{_j}$ is the atomic mass and $\omega$ is the phonon frequency. The results of these calculations are reported in Supplementary Note 7.\\

\subsection*{\textit{Ab initio} calculations - Indirect excitations}

Anatase TiO$_2$ is an indirect bandgap material with a minimum indirect gap amounting to 3.61 eV, according to our calculations without ZPR corrections. This gap is smaller than the optical gap we obtained at the BSE level of theory (3.76 eV). As discussed in the main text, exciton I is bound with respect to the direct gap. Despite this, the exciton could also be considered as resonant with respect to all phonon-mediated indirect transitions. In the frozen-phonon calculations for a single TiO$_2$ unit cell, the BSE includes only coupling of direct electron-hole transitions with phonons at the $\Gamma$ point, and hence, possible effects originating from the indirect nature of the material would not be accounted for. A way to incorporate those effects is to perform BSE calculations for a large TiO$_2$ supercell, where both the indirect- and direct gap are folded into the $\tilde{\Gamma}$ point of the supercell. In such a calculation, frozen atom displacements can couple electron and hole states with different $k$ values in the original sampling of the first BZ via phonons with nonzero \textit{q}-vectors. The results of these calculations are reported in Supplementary Note 8. We have recently become aware of Ref. 68, in which a one-shot method to compute the indirect contributions to the absorption tail using supercell methods has been proposed.\\

\noindent \textbf{Acknowledgments:} We thank Prof. Fabrizio Carbone for useful discussions, Dr. Luca Moreschini for the technical support during the ARPES measurements and the team of the Electronic Structure Factory endstation on beamline 7.0.1 at the Advanced Light Source for the doping dependence of the ARPES measurements. We acknowledge financial support from the Swiss NSF via the NCCR:MUST and the contracts No. 206021\_157773, 20020\_153660 and 407040\_154056 (PNR 70), the European Research Council Advanced Grants H2020 ERCEA 695197 DYNAMOX and QSpec-NewMat (ERC-2015-AdG-694097), Spanish Grant FIS2013-46159-C3-1-P, Grupos Consolidados del Gobierno Vasco (IT578-13), COST Actions CM1204 (XLIC), MP1306 (EUSpec), European Union’s H2020 program under GA no.676580 (NOMAD) and the Austrian Science Fund (FWF P25739-N27). M.P. acknowledges the EC for RISE project no CoExAN GA644076 and Cineca for computational resources allocated under Iscra-C initiative. The Advanced Light Source is supported by the Director, Office of Science, Office of Basic Energy Sciences, of the U.S. Department of Energy under Contract No. DE-AC02-05CH11231.

\newpage

\begin{figure}[h!]
	\includegraphics[width=\textwidth]{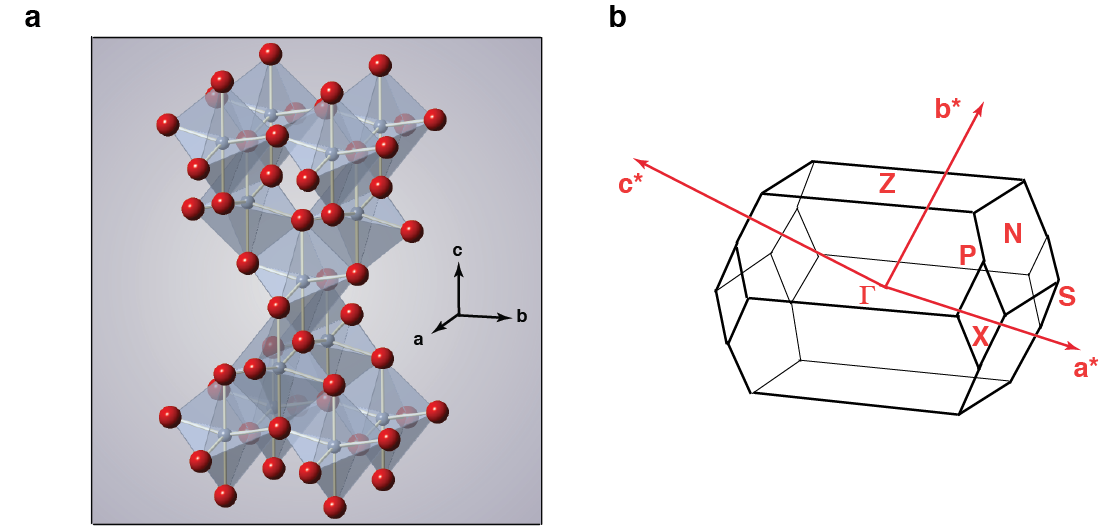}
\caption{\textbf{Anatase TiO$_2$ crystal structure and Brillouin zone.\\}\textbf{(a)} Crystallographic structure of anatase TiO$_2$ with highlighted TiO$_6$ polyhedra. Blue atoms represent titanium, red atoms represent oxygen. \textbf{(b)} Representation of the 3D BZ of anatase TiO$_2$.}
\end{figure}
\newpage

\begin{figure}[h!]
	\includegraphics[width=0.8\textwidth]{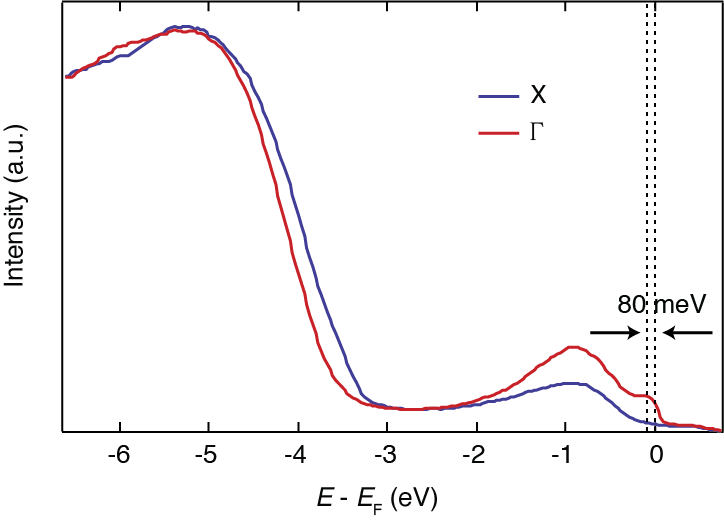}
\caption{\textbf{ARPES energy distribution curves.\\}Energy distribution curves at the X (blue curve) and $\Gamma$ (red curve) points of the BZ for a crystal at 20 K doped with an excess electron density $n$ = 2 $\times$ 10$^{19}$ cm$^{-3}$. The spectra exhibit a feature at -1 eV, corresponding to the in-gap oxygen defect states. The curve at $\Gamma$ exhibits spectral weight at $E_\mathrm{F}$, which lies 80 meV above the conduction quasiparticle band. The spectrum is referenced to $E_\mathrm{F}$.}
\end{figure}
\newpage

\begin{figure}[h!]
	\includegraphics[width=\textwidth]{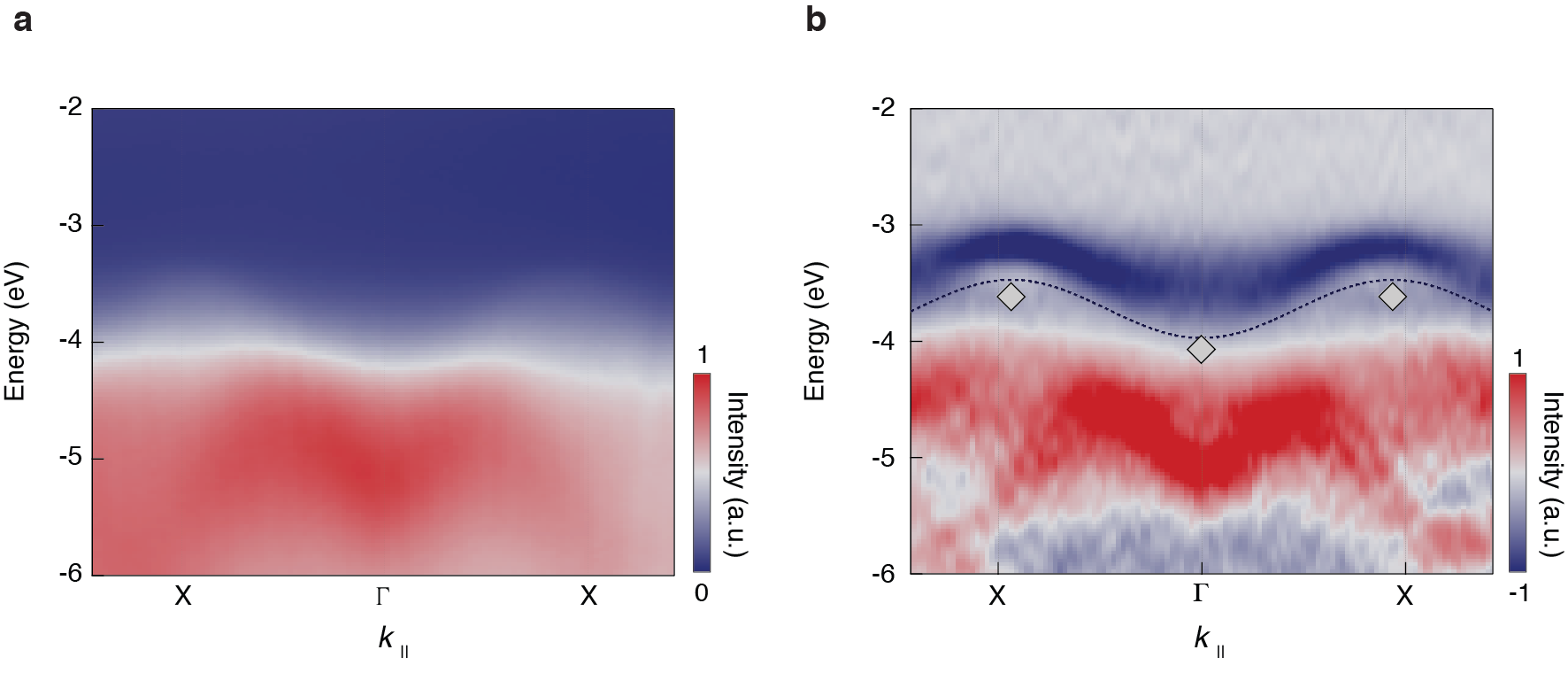}
\caption{\textbf{ARPES data of a single crystal of anatase TiO$_2$.\\}\textbf{(a)} ARPES energy vs. momentum intensity map for a crystal doped with an excess electron density $n$ = 2 $\times$ 10$^{19}$ cm$^{-3}$ at 20 K. \textbf{(b)} Second derivative ARPES data of the electronic structure at the top of the VB between $\Gamma$ and X. Dashed blue lines are added as a guide to the eye. The spectrum is referenced to the bottom of the CB at $\Gamma$. The intensity is indicated by a linear colour scale, as reported in the colour bar.}
\end{figure}
\newpage

\begin{figure}[h!]
\includegraphics[width=\textwidth]{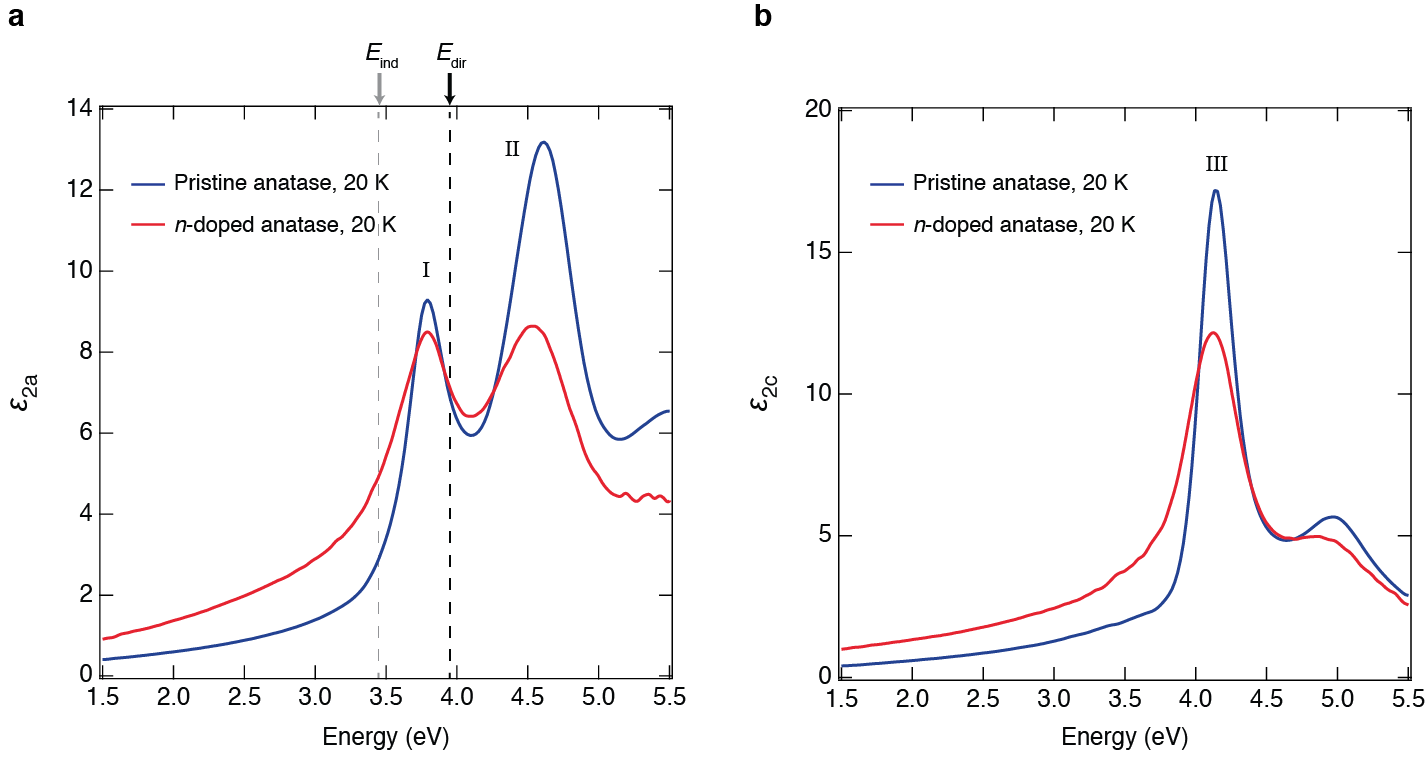}
\caption{\textbf{Optical spectra of anatase TiO$_2$ single crystals.\\} Imaginary part of the dielectric function at 20 K with the electric field polarized along \textbf{(a)} the a-axis (\textbf{E} $\perp$ c) and \textbf{(b)} the c-axis (\textbf{E} $\parallel$ c). The experimental data measured by SE on a pristine ($n$ $\sim$ 0 cm$^{-3}$) anatase TiO$_2$ single crystal are reported in blue, while those obtained on a highly $n$-doped single crystal ($n$ = 2 $\times$ 10$^{19}$ cm$^{-3}$) in red. The quasiparticle indirect gap $E\mathrm{_{ind}}$ = 3.47 eV and direct gap $E\mathrm{_{dir}}$ = 3.97 eV, as estimated by ARPES, are indicated by dashed grey and black vertical lines, respectively.}
\end{figure}
\newpage

\begin{figure}[h!]
	\includegraphics[width=0.7\textwidth]{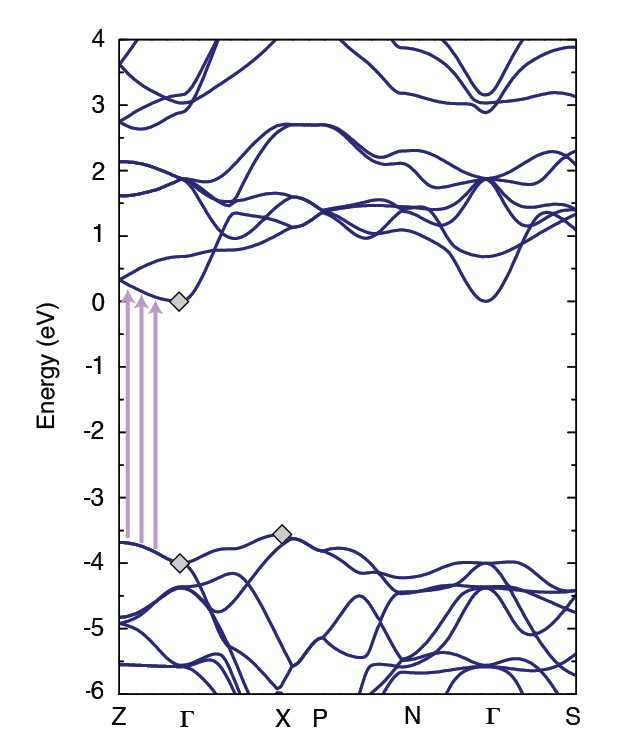}
\caption{\textbf{Calculated electronic structure.\\} Electronic band structure of pristine anatase TiO$_2$ in the first BZ. The lines indicate the DFT calculations corrected by the GW values. Grey diamond dots indicate the values obtained with the GW corrections at the $\Gamma$ and X points, while the values along the high symmetry directions are obtained by correcting with linearly interpolated values.}
\label{fig:FigS4}
\end{figure}
\newpage

\begin{figure}[h!]
	\centerline{\includegraphics[width=\textwidth]{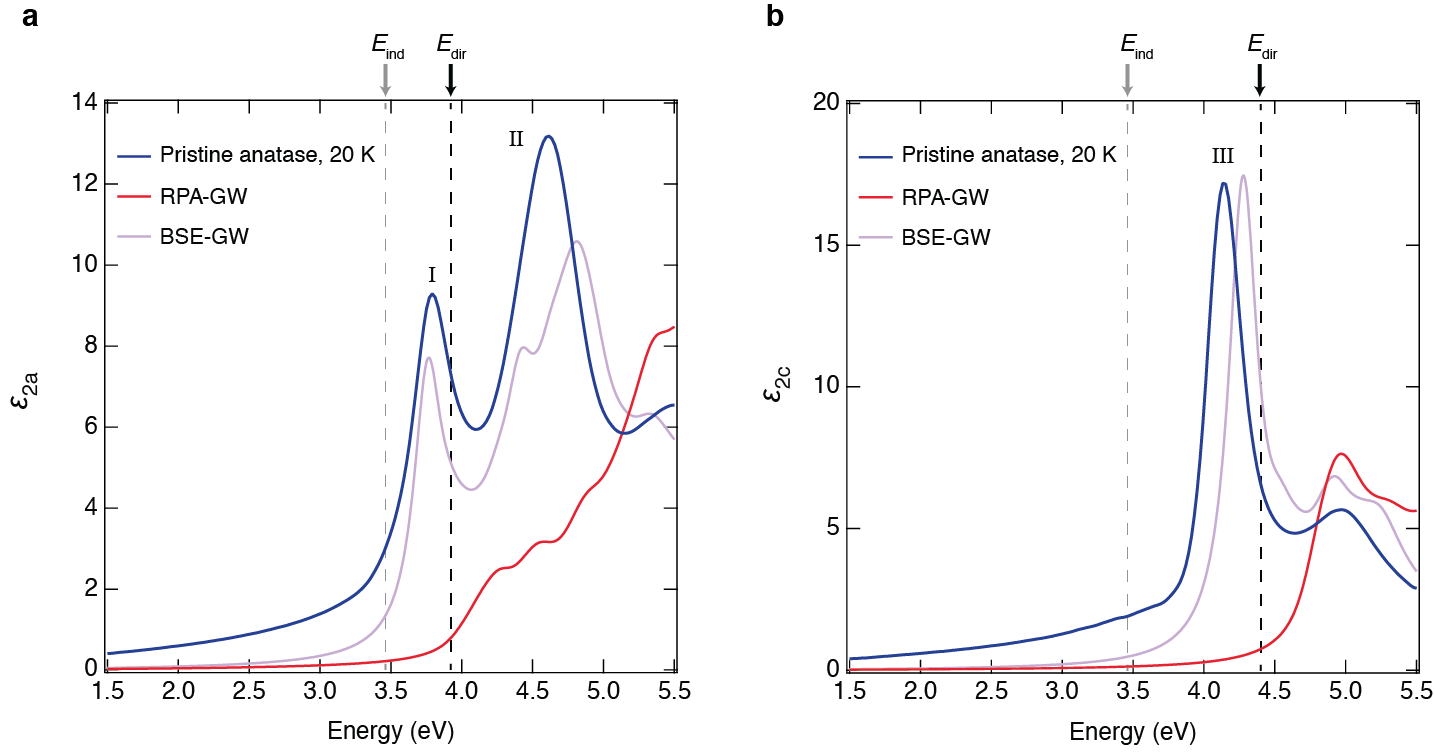}}
\caption{\textbf{Calculated optical spectra of pristine anatase TiO$_2$.\\} \textbf{(a, b)} Comparison between the spectra measured at 20 K on the pristine anatase TiO$_2$ single crystal and those obtained from frozen-lattice \textit{ab initio} calculations for pristine anatase TiO$_2$. The experimental data are reported in blue, the calculated spectra in the RPA-GW scheme in red and the calculated spectra in the BSE-GW scheme in violet. The quasiparticle indirect gap $E\mathrm{_{ind}}$ = 3.46 eV is indicated by a dashed grey vertical line; the direct gaps $E\mathrm{_{dir}}$ = 3.92 eV (at the $\Gamma$ point) for \textbf{E} $\perp$ c and $E\mathrm{_{dir}}$ = 4.40 eV (coincident with the onset of the RPA-GW) for \textbf{E} $\parallel$ c are indicated by dashed black vertical lines in Fig. \textbf{a} and \textbf{b}, respectively.}
\end{figure}
\newpage

\begin{figure}[h!]
	\includegraphics[width=\textwidth]{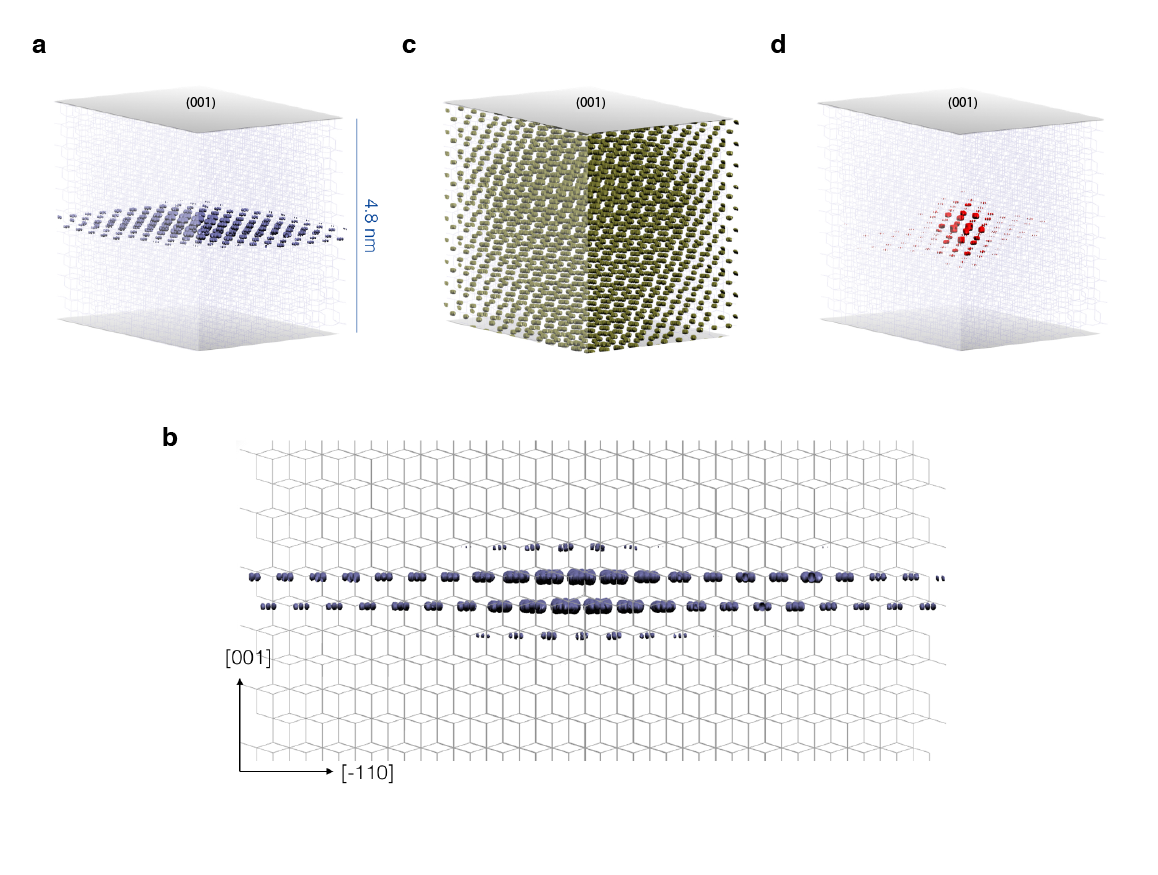}
\caption{\textbf{Wavefunctions of the fundamental charge excitations in anatase TiO$_2$.\\} Isosurface representation of the electronic configuration when the hole of the considered excitonic pair is localized close to one oxygen atom. The coloured region represents the excitonic squared modulus wavefunction. \textbf{(a)} Bound exciton I at 3.76 eV. \textbf{(b)} Side-view of the bound exciton I at 3.76 eV. \textbf{(c)} Resonance II at 4.37 eV. \textbf{(d)} Bound exciton III at 4.28 eV.}
\end{figure}
\newpage

\begin{figure}[h!]
	\centerline{\includegraphics[width=0.55\textwidth]{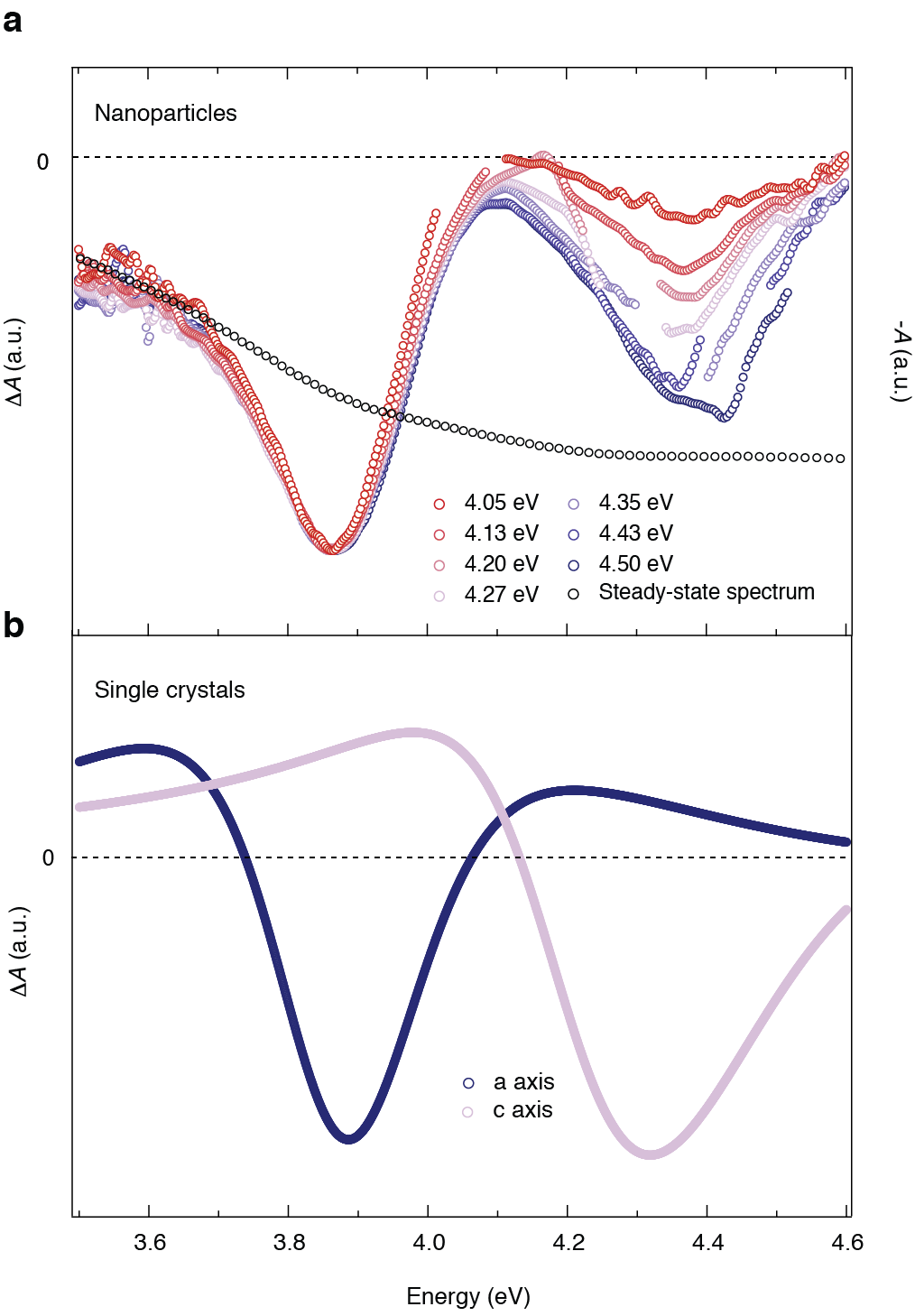}}
\caption{\textbf{Ultrafast 2D UV spectroscopy of anatase TiO$_2$ NPs and single crystals.\\} \textbf{(a)} Normalized transient absorption ($\Delta A$) spectra of RT colloidal solution of anatase TiO$_2$ NPs at a fixed time-delay of 1 ps and for different pump photon energies (indicated in the figure). Each trace is normalized with respect to the minimum of the main feature at 3.88 eV. For comparison, the black trace shows the inverted steady-state absorption spectrum. \textbf{(b)} $\Delta A$ spectrum of RT anatase TiO$_2$ single crystals along the a- and c-axis at a fixed time delay of 1 ps. For this experiment, the pump photon energy is 4.40 eV.}
\end{figure}
\newpage

\label{Supplementary Information}

\newpage
\renewcommand{\thesection}{S\arabic{section}}  
\renewcommand{\thetable}{S\arabic{table}}  
\renewcommand{\thefigure}{S\arabic{figure}} 
\renewcommand\Im{\operatorname{\mathfrak{Im}}}
\setcounter{figure}{0}

\section*{Additional experimental results}

\subsection*{Supplementary Note 1: Steady-state spectroscopic ellipsometry}
\label{Results_Ellipsometry}

\noindent\textbf{Comparison with previous experiments.} Our spectroscopic ellipsometry (SE) measurements are performed on two classes of (010)-oriented single crystals, namely pristine anatase ($n$ $\sim$ 0 cm$^{-3}$) and $n$-doped anatase ($n$ = $2$ $\times$ 10$^{19}$ cm$^{-3}$). The reflectivity response derived from SE on the pristine (violet lines) and $n$-doped (blue lines) single crystals is plotted in Supplementary Figs. \ref{fig:FigS2}a-b together with the data of Ref. 1 (red lines). The reflectivities for light polarized along the a- and c-axis of the crystals are shown in Supplementary Figs. \ref{fig:FigS2}a and \ref{fig:FigS2}b, respectively. To establish a direct comparison with the normal-incidence reflectivity data of Ref. 1, all reported traces have been measured at 100 K. We observe the overall agreement of our results with the ones reported in literature, with slight shifts in the reflectivity peak energies. However, given the sharpness of the reflectivity lineshapes, we can also deduce that the quality of our pristine crystal is much higher than the one employed in Ref. 1. Indeed, the reflectivity spectrum of the latter is closer to the response measured on our $n$-doped crystal, which has a large percentage of oxygen vacancies. Furthermore, we note that the two different energy scales used in Ref. 1 to display the data of the dielectric function for \textbf{E} $\perp$ c and \textbf{E} $\parallel$ c has led to an error of digitalization by early computational works, which has propagated in the literature \cite{ref:chiodo, ref:kang, ref:landmann, ref:lawler} on anatase TiO$_2$.

\noindent Supplementary Figs. \ref{fig:FigS2}c-d display the normalized absorption spectra of our anatase TiO$_2$ single crystals at 300 K. The a- and c-axis responses are respectively shown in Supplementary Figs. \ref{fig:FigS2}c and \ref{fig:FigS2}d. Here, our interest is to reveal the robustness of the excitonic peaks even at room temperature (RT) and to highlight the role played by the presence of oxygen vacancies in the doped sample. Hence, for this purpose, the data are normalized with respect to the lowest direct exciton peaks. We observe that in both the a- and c-axis absorption spectra the exciton peak energies are not renormalized by the presence of oxygen vacancies, while the linewidth becomes broader in the $n$-doped crystal.

\noindent Figures \ref{fig:Epsilon_KK}a and \ref{fig:Epsilon_KK}b display $\epsilon_1(\omega)$ and $\epsilon_2(\omega)$ calculated by applying the Kramers-Kronig (KK) analysis on the reflectivity spectrum at RT. This technique represents a valuable test to evaluate the consistency with our original $\epsilon_1(\omega)$ and $\epsilon_2(\omega)$ values. We observe that the effect of the KK transformation on both the a- and c-axis $\epsilon_2(\omega)$ is to modify the lineshape and the intensity of the peaks and to change the spectral weight in the Urbach tail. This behaviour shows that the KK analysis, even performed on the broadest possible spectral range, is not a precise approach to treat the normal-incidence reflectivity data.\\

\noindent\textbf{Many-body effects.} In the case of the $n$-doped ($n$ = 2 $\times$ 10$^{19}$ cm$^{-3}$) anatase TiO$_2$ crystals, the presence of an excess electron density at the $\Gamma$ point of the CB may give rise to a variety of effects having a profound impact on both the single-particle and the two-particle excitation spectra. According to their origin, these effects can be distinguished between those involving single-particle states (phase-space filling) and those that can be attributed to many-body interactions among the doped carriers (long-range Coulomb screening and band gap renormalization, BGR). Phase-space filling arises because of the Pauli exclusion principle, which applies to the electrons and holes constituting the excitons. This produces a finite exclusion volume in phase-space for each exciton. As a consequence, the VB-to-CB transition probability is reduced, which is evidenced by a reduction of oscillator strength of the excitonic transition in the optical spectra.
On the other hand, many-body interactions alter the exciton energy and composition by affecting the underlying electronic states. This occurs via the direct and exchange Coulomb interactions of electrons and holes, which at high plasma density provide additional screening channels. First, long-range free-carrier screening modifies the exciton Coulomb potential $\phi \sim \frac{e}{4\pi\epsilon r}$ through a multiplicative factor e$^{-r/\lambda}$, where $\lambda$ is the Debye length. Substantial screening occurs when the Debye length and the exciton radius become comparable. Second, the excess electrons also lead to BGR, \textit{i.e.} a density-dependent shrinkage of the quasiparticle gap due to electron and hole self-energy corrections. As a result, the enhancement of the electronic screening leads to the simultaneous renormalization of both the exciton $E\mathrm{_B}$ and the electronic gap. In this scenario, the exciton absorption energy in the SE data is determined by the combination of the weakened Coulomb interaction and the BGR, the former inducing a blueshift and the latter a redshift of the exciton peak. The quantitative details of this compensation depend on both material and dimensionality but this arises as a general effect in many standard bulk semiconductors~\cite{ref:reynolds} and nanostructures~\cite{ref:wegscheider, ref:ambigapathy}. In the low-density limit, they have been shown to cancel to first order~\cite{ref:dassarma}. Relying on this discussion, we can provide a deeper interpretation of the changes in the low-temperature dielectric function of anatase TiO$_2$ upon electron doping (Fig. 4). The redistribution of spectral-weight can be attributed to the combination of phase-space filling and defect-induced in-gap absorption. The enhanced broadening of the exciton lineshapes can be directly associated with a modification of the exciton lifetime, due to the combination of long-range Coulomb screening and exciton-defect scattering. Finally, the insensitivity of the exciton energies to the effective doping level suggests that the long-range Coulomb screening and the BGR perfectly compensate each other even at high doping levels. This idea is reinforced by our \textit{ab initio} calculations (see Supplementary Note 6), which reveal that doping-induced many-body effects in anatase TiO$_2$ play a marginal role even at high carrier densities.

\subsection*{Supplementary Note 2: Ultrafast transient reflectivity of single crystals}

Femtosecond transient reflectivity experiments are performed on three different classes of (001)- and (010)-oriented single crystals, with $n$ $\sim$ 0 cm$^{-3}$, $n$ = 2 $\times$ 10$^{17}$ cm$^{-3}$ and $n$ = 2 $\times$ 10$^{19}$ cm$^{-3}$. 

\noindent In a first experiment, the $\Delta R/R$ signal is monitored in a broadband UV range (3.75 - 4.35 eV) for the three classes of samples along the ab planes. Both pump and probe polarizations are set in a parallel configuration. An isotropic optical response is found when the (001)-oriented crystals are rotated about the c-axis. The pump energy is 4.10 eV, in order to selectively perturb the spectral region above the first excitonic peak. Supplementary Figure \ref{fig:FigS4}a compares the transient spectrum obtained from the three specimens at a fixed time-delay of 6 ps. The $\Delta R/R$ spectrum shows an inversion point around 3.96 eV. The inflection that is present around 4.10 eV in the signal obtained from the sample with $n$ = 2 $\times$ 10$^{17}$ cm$^{-3}$ is an artefact produced by the scattering of the pump beam in our spectrometer. A similar response can be found for all the temporal delays between pump and probe up to 1 ns. The main difference is displayed by the intensity of the signal at long time delays, since it depends on the rate of the carrier recombination process. Supplementary Figure \ref{fig:FigS4}b shows three temporal traces up to 100 ps probed around 3.82 eV, in which this effect is clearly visible. The decay of the nonequilibrium signal in the strongly $n$-doped crystal is faster than the other responses, since the increased density of in-gap states facilitates charge carrier recombination across the bandgap. The detailed analysis of the kinetics will be subject of a separate publication.

A second set of experiments is performed on (001)- and (010)-oriented $n$-doped ($n$ = 2 $\times$ 10$^{19}$ cm$^{-3}$) single crystals of anatase TiO$_2$, in order to access the anisotropic dynamics along the a- and c-axis. Supplementary Figures \ref{fig:Composite}a and \ref{fig:Composite}c display the $\Delta R/R$ maps of the a- and c-axis response for the (010)-oriented single crystal as a function of the probe photon energy and of the time delay between pump and probe. Although the kinetics are measured up to 1 ns, the two maps are displayed up to 10 ps. The a-axis response is measured upon photoexcitation at 4.40 eV, with the broadband probe covering the range 3.70 - 4.65 eV. These results can be also reproduced by polarizing the beams along the a-axis of the (001)-oriented samples. The higher-energy region of the spectrum does not evolve in time, remaining unaffected by the photoexcitation process. The c-axis response (Supplementary Fig. \ref{fig:Composite}c) is obtained through the measurement of the (010)-oriented single crystal with a pump and probe polarizations set along the c-axis. Also in this case the pump energy is at 4.40 eV, but the probed range is shifted to 4.05 - 4.80 eV. The transient spectrum of Supplementary Fig. \ref{fig:Composite}d strongly differs from the in-plane one, consisting of a negative contribution set around 4.28 eV and a tail extending to 4.60 eV in the high-energy range.

In a final experiment, both the pump (at 4.40 eV) and the probe (between 3.70 and 4.60 eV) beams are polarized along the c-axis of the $n$ = 2 $\times$ 10$^{19}$ cm$^{-3}$ single crystal. The resulting $\Delta R/R$ map is displayed in Supplementary Fig. \ref{fig:TransientReflectivity_caxis}. This demonstrates the absence of c-axis spectral features at low energies and confirms the finding of the pump-probe experiment along the c-axis shown in Supplementary Figs. \ref{fig:Composite}c-d. The difference in terms of intensity with respect to those measurements has to be attributed to the reduced pump intensity (of $\sim$ 1/3) due to constraints in the generation of the broadband probe beam.

\newpage

\section*{Data analysis and theory}

\subsection*{Supplementary Note 3: Extracting $\Delta A$ from $\Delta R/R$ in single crystals}

In our ultrafast measurements, we probe $\Delta R/R$ of anatase \ce{TiO2} single crystals and $\Delta A$ of a colloidal solution of NPs. Thus, it is useful to analyze the time-resolved dynamics of the single crystals in terms of their transient absorption, in order to establish a link with the data on NPs. However, we recall that $\Delta R/R$ has a non-trivial relationship with both $\Delta\epsilon_1$ and $\Delta\epsilon_2$ in the probed spectral range. Indeed, in the UV, the real and imaginary part of the dielectric function have rather similar absolute values. For this reason, the optical reflectivity is equally sensitive to the reactive and the absorptive components of the dielectric function.

Hence, in order to calculate the pump-induced evolution  of the $\Delta A$ from the $\Delta R/R$ data, we proceed as follows. We model the steady-state SE data using a set of Lorentz oscillators, we calculate the equilibrium reflectivity ($R\mathrm{_{0}}$), and we fit the measured transient reflectance $R\mathrm{_{exp}}(t)$/$R\mathrm{_{exp}}$ with a differential model ($R(t)$ - $R\mathrm{_0}$)/$R\mathrm{_{0}}$, where $R(t)$ is a model for the perturbed reflectivity obtained by variation of the parameters used to fit the equilibrium data as a function of the pump-probe delay $t$. We adopt this approach to treat our $\Delta R/R$ data to avoid possible systematic errors that can be produced by the typical analysis through KK transformations. The SE spectra were fitted using a dielectric function of the form
\begin{equation}
\epsilon(\omega) = \epsilon_\infty + \sum_\mathrm{j}\frac{\omega\mathrm{^2_{pj}}}{\omega\mathrm{^2_{0i}}-\omega^2-i\Gamma\mathrm{_j}\omega}
\end{equation}

\noindent where $\epsilon_\infty$ is the high-frequency dielectric constant, and $\omega\mathrm{^2_{pj}}$, $\omega\mathrm{_{0j}^2}$, $\Gamma\mathrm{_j}$ are, respectively, the plasma frequency, the transverse frequency and the linewidth (scattering rate) of the \textit{j}-th Lorentz oscillator. The absorbance is then given by: $A(\omega)$ = $\omega$ $\operatorname{Im}\sqrt\epsilon$.

\noindent For the fitting of the transient data, the Lorentz oscillators in our experimental range are allowed to change in order to reproduce the dynamical reflectivity. This procedure enables to extract the transient dielectric function $\Delta\epsilon(\omega, t)=\Delta\epsilon_1(\omega, t)+i\Delta\epsilon_2(\omega, t)$ and finally leads to the evaluation of the $\Delta A$ for the single crystals.

\subsection*{Supplementary Note 4: \textit{Ab initio} calculations - Frozen-lattice results}
\label{AbInitioFrozen}

The calculated GW direct bandgap at the $\Gamma$ point is 4.07 eV, at Z it is 4.13 eV, and the indirect bandgap (between $\Gamma$ and a $k$-point close to X) is 3.61 eV. The bandgap at the middle point of the $\Gamma$-Z line is 3.96 eV. These values have been converged up to 5 meV, and the two codes used for the calculations give the same results, despite the use of a different plasmon pole models for the frequency integration in the GW method. The present fully-converged minimum GW quasiparticle correction amounts to 1.40 eV, which is smaller than the value of 1.69 eV from Ref. 2 (the difference comes from the smaller number of bands and $k$-points used in Ref. 2), highlighting the careful and exhaustive convergence evaluation done in the present work.

The symmetry-line along $\Gamma$-Z shows nearly parallel dispersion curves for the CB and VB edges. This peculiar shape of electronic states along $\Gamma$-Z plays a fundamental role in the optical properties of the material, as it dictates the nature and binding of the lowest excitons in anatase TiO$_2$ (see below). The nearly parallel dispersion observed in the theoretical bandgap allows us to use the direct gap at $\Gamma$ as a very good approximation to estimate the bound direct nature of the exciton to be compared with the experimental data. Due to the band structure shape along the $\Gamma$-Z high-symmetry direction, we also underline that an accurate $k$-point sampling is especially critical for the quality of the optical spectra, since the main excitons are built up from optical transitions with contributions from a small region of the BZ. The CB and VB in this region display a wormlike shape aligned along the $\Gamma$-Z direction.

In Supplementary Fig. \ref{fig:BSECalculations}, we compare the results obtained with the GW implementations in BerkeleyGW and Yambo codes obtained at the same level of accuracy. Yambo calculations were performed using a 12$\times$12$\times$12 unshifted grid whereas a randomly shifted grid of 8$\times$8$\times$8 $k$-points was employed in the BerkeleyGW calculations. Thus, both approaches employ roughly 500 $k$-points. This shows the equivalence between the two implementations. To get the fully-converged optical spectra shown in the main text, a denser grid is required. Most importantly, the main effect of the stringent convergence obtained here with respect to $k$-points and number of bands is given by the shape of exciton I (see Supplementary Fig. 11). This charge excitation, split in two small peaks at low convergence \cite{ref:chiodo, ref:kang} (or a main peak with a shoulder), becomes a unique, uniform peak, similar to the one observed in the experiment (dark red curve in Supplementary Fig. 11). The fine $k$-sampling is needed, since the main optical transitions contributing to exciton I comes from the $\Gamma$-Z line, with bands almost parallel and flat. The two-dimensional exciton I for \textbf{E} $\perp$ c (at 3.76 eV) has indeed a major contribution from the transition from the top of the VB to the bottom of the CB at the middle point in the $\Gamma$-Z line. To a lesser extent, significant contributions come from the $k$-points lying along the $\Gamma$-Z line and close to it in every direction. The contribution increases gradually when approaching the aforementioned $\Gamma$-Z middle point. Even if the GW direct electronic band gap of 3.96 eV, that is located at the middle point along the $\Gamma$-Z line, is used as a reference energy to calculate the exciton binding energy ($E\mathrm{_B}$), the bound nature of exciton I is still confirmed (with $E\mathrm{_B}$ = 50 meV). A phenomenological Lorentzian broadening of 0.12 eV was applied to reproduce the experimental spectra (see Supplementary Fig. 12 for a comparison of the measured spectrum with the bare BSE eigenvalues, showing that mainly one eigenvalue is contributing to the exciton peak I and that the measured lifetime is not of electronic origin but is due to the strong electron-phonon coupling in this material). 

The energy, shape and reciprocal space contributions for peak II highlights its bulk-resonance character, most evident as its offset coincides with the RPA-GW absorption rise. On the other hand, we found that exciton III for \textbf{E} $\parallel$ c (at 4.28 eV) is of a more complex nature, with a mixed contribution of bound excitons and delocalized resonant transitions. The former have the dominant character and the contributing transitions are found throughout the BZ, while the main contribution for the latter is concentrated in few points close to the $\Gamma$-Z line in the region around Z. The energy and shape of exciton III confirm the analysis of a mixed bulk resonance and localized character, as the continuum onset in RPA-GW appears to undergo an intensity enhancement. The mixed bulk-localized nature makes it less straightforward to estimate its $E\mathrm{_B}$. Assuming the RPA-GW onset at 4.40 eV for \textbf{E} $\parallel$ c as the reference energy for evaluating $E\mathrm{_B}$, we estimate $E\mathrm{_B}$ $\sim$150 meV.

The slight shift of the calculated exciton III with respect to our experimental value (0.1 eV) is also investigated in detail. Increasing the number of $k$-points and bands does not allow to obtain a match to experiment as good as for exciton I. We can also exclude possible effects of anisotropic screening, as increasing the parameters of the local field effects and separating the screening along the a- and c-axis components do not lead to significant changes in the spectrum. The peak maximum seems instead to be related, in a nonlinear manner, to the lattice constants. The a-axis lattice constant from \textit{ab initio} optimization is in excellent agreement with experimental data (3.79 \AA~vs 3.78 \AA), while the c-axis lattice constant is slightly (1\%) overestimated. However, when using the experimental lattice parameters, the position of peak III is blueshifted by 0.20 eV from that obtained with the PBE parameters, thus worsening the agreement with experiment. 

Finally, the presence of dark excitons in anatase TiO$_2$ at energies below the bright exciton I at 3.76 eV is ruled out by our spin-resolved optical BSE calculations. Indeed, the calculated lowest exciton in anatase TiO$_2$ is a singlet and it is optically active. This does not apply to the rutile phase of TiO$_2$ \cite{ref:chiodo}, where the lowest exciton is an optically dark triplet state.

\subsection*{Supplementary Note 5: \textit{Ab initio} calculations - Comparison with previous studies}
\label{AbInitio_Comparison}

The computational data presented in this paper are novel and conclusive in many respects, despite being performed at the same level of theory (DFT + GW + BSE) as in some previous reports \cite{ref:chiodo, ref:kang, ref:landmann, ref:lawler}. First of all, this is the first available comparison between the theoretical and the experimental electronic gap for pristine and doped anatase TiO$_2$. As the experimental gap at the $\Gamma$ point of the BZ is measured here for the first time, in the past there has been a high degree of uncertainty concerning the computational results. The reported direct electronic gaps at $\Gamma$ ranged from 4.14~eV (PBE-G$_0$W$_0$) \cite{ref:kang} to 3.78~eV (PBE-G$_0$W$_0$) \cite{ref:landmann} and 4.29 eV (PBE-G$_0$W$_0$) \cite{ref:chiodo}. The use of methods beyond GGA-G$_0$W$_0$ provided even larger values (4.05 eV, G$_0$W$_0$ on top of hybrid functional HSE06 for the indirect gap \cite{ref:landmann}, larger than our 3.61 eV; 5.28~eV, self-consistent GW different flavours \cite{ref:g_kang}).

\noindent Moreover, the GW calculations presented here go beyond the ones reported in previous works, as they address the combined effects of electron doping, temperature-induced lattice expansion and electron-phonon coupling on the direct gap at the $\Gamma$ point of the BZ. Such investigations rule out the role of the BGR in the description of the quasiparticle gap and enable the identification of electron-phonon coupling as the main source of renormalization of the quasiparticle gap.

\noindent A second important point concerns the comparison of the refined theoretical data here reported with SE data. Indeed, the previous zero-temperature computational calculations relied on a comparison with the dielectric function at 100 K of Ref. 1, which was extracted via a KK analysis from normal-incidence reflectivity data. We note again that the two different energy scales used in Ref. 1 to display the data of the dielectric function for \textbf{E} $\perp$ c and \textbf{E} $\parallel$ c has led to an error of digitalization by early computational works, which has propagated in the literature \cite{ref:chiodo, ref:kang, ref:landmann, ref:lawler} on anatase TiO$_2$. Here, we correctly compare our new calculations with the low-temperature dielectric function of the material, measured directly via SE and not extracted through a KK analysis. In this way, we clarify the precise peak positions and shape in the optical absorption for the material. 

\noindent Besides the higher convergence achieved in comparison with previous works, our evaluation of the optical spectra presents novel results, since we included doping, electron-phonon coupling and the role of indirect transitions, and discussed their effects on the energy and shape of the optical peaks (see the following sections).

\subsection*{Supplementary Note 6: \textit{Ab initio} calculations - Doped anatase TiO$_2$}
\label{AbInitio_Doping}

Within the same theoretical framework used for pristine anatase TiO$_2$, we perform calculations for the case of uniformly doped anatase TiO$_2$, to verify computationally that the influence of doping on both the electronic gap and optical response can be disregarded. Here, we show the results for two cases of uniform excess electron density $n$ = 10$^{19}$ cm$^{-3}$ and  $n$ = 10$^{20}$ cm$^{-3}$. 

\noindent The calculated GW gap (both  direct and indirect gap) are similar to the pristine anatase TiO$_2$ case, with an increase of 1 meV for the doping of $n$ = 10$^{19}$ cm$^{-3}$ and of 17 meV for $n$ = 10$^{20}$ cm$^{-3}$. These results complement the experimental ARPES data, demonstrating that the electronic gap from doped samples is a very good value to describe also the gap of pristine anatase TiO$_2$. 
In the presence of doping, two competing effects contribute to changing the electronic gap of a semiconductor, with either a redshift or blueshift depending on which effect dominates. The CB filling is responsible for the blueshift, while the change in the long-range Coulomb screening (becoming slightly metallic) is responsible for the redshift. In anatase TiO$_2$, the dominant effect for the considered doping ranges is the CB filling. For the doping values relevant in our measurements, the correction is well below the computational error, therefore, there is no detectable effect of doping in the electronic band gap.
This is supported by the calculation of the optical response, as the position and shape of peak I in $\epsilon_\mathrm{{2a}}$ (Supplementary Fig. \ref{fig:DopingDependence}) do not change for $n$ = 10$^{19}$ cm$^{-3}$. These results justify the strategy, followed in the main text, to experimentally estimate the exciton $E\mathrm{_B}$ from the ARPES measurements in the doped samples.

\subsection*{Supplementary Note 7: \textit{Ab initio} calculations - Electron-phonon and temperature effects}
\label{AbInitio_Temperature}

Our calculations taking into account the role of the electron-phonon coupling reveal a GW bandgap increase of 60 meV (in the case of the $\mathrm{E_{u}}$ mode) and 80 meV (in the case of the $\mathrm{A_{2u}}$ mode) at 300 K, compared to the zero temperature value. We correct this value by also considering the lattice expansion effect. By using the thermal expansion coefficient in Ref. 12, we determine that the a and c lattice parameters of anatase TiO$_2$ increase in 0.1 $\%$ and 0.3 $\%$, respectively, from 0 to 300 K. The inclusion of both the phonon-induced and thermal expansion-induced effects leads to a net blueshift of the bandgap of about 30 meV (in the case of the $\mathrm{E_{u}}$ mode) and 50 meV (in the case of the $\mathrm{A_{2u}}$ mode) from zero temperature to 300 K. A similar trend was recently reported for rutile TiO$_2$, where the electronic gap (evaluated within the thermal lines method for electron-phonon coupling) has a non-monotonic behavior with temperature \cite{ref:monserrat}. Additionally, we solve the BSE on top of the temperature-corrected GW and find a net blueshift of roughly 80 meV (in the case of the $\mathrm{E_{u}}$ mode) 70 meV (in the case of the $\mathrm{A_{2u}}$ mode) at 300 K, which is in line with our SE measurements (blueshift of 40 meV from 20 K to 300 K). 

\subsection*{Supplementary Note 8: \textit{Ab initio} calculations - Direct exciton in an indirect band gap material}

We consider a supercell composed of 3$\times$3$\times$2 conventional unit cells (12 atoms) which leads to a total of 216 atoms in the supercell. This implies the inclusion of 648 phonons. Although the employed number of phonons is still limited, it provides a first approximation of the effect of the indirect gap in the renormalization of the excitonic peak. We perform two molecular dynamics simulation runs at temperatures of 20 K and 300 K. The molecular dynamics runs were carried out using a Nos\'e-Hoover chain thermostat. A total of 5 snapshots were randomly chosen in the interval between 5 ps and 10 ps of the run for each temperature. To investigate if the position of the excitonic peak changes when accounting for the indirect nature of the material, we perform similar calculations for the primitive unit cell of anatase TiO$_2$ at the same level of theory and convergence. We obtain a negligible blueshift of 30 meV, which indicates that the indirect bandgap nature of anatase TiO$_2$ does not play a significant role in the exciton properties, beyond adding an Urbach tail at the lower energy side of the peak.

\clearpage
\newpage 

\renewcommand{\refname}{Additional References}

\newpage
\clearpage

\begin{figure}[h!]
	\begin{center}
		\includegraphics[width=0.5\textwidth]{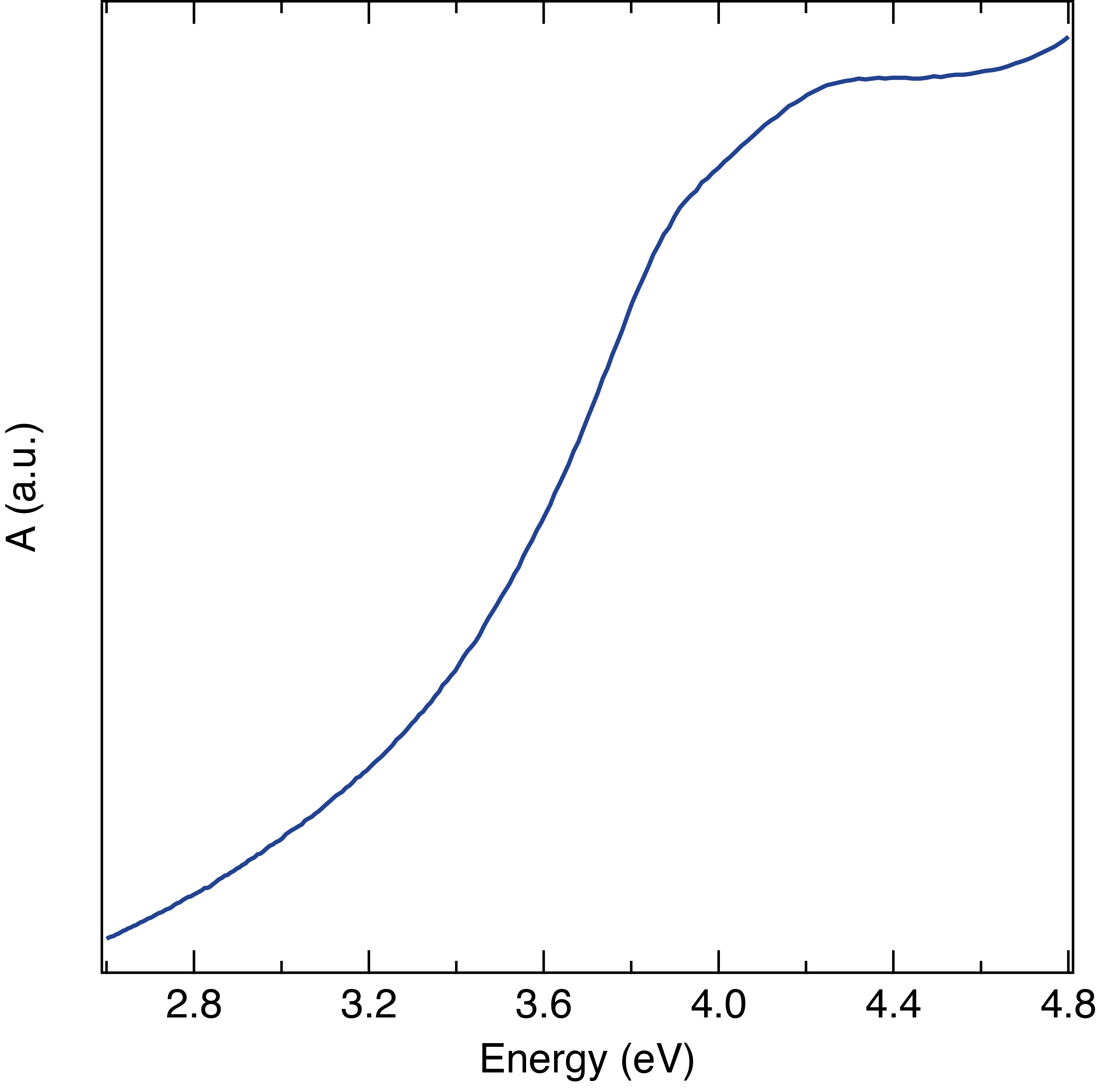}
		\caption{\textbf{Steady-state absorption spectrum of anatase TiO$_2$ NPs.}\\ Room temperature steady-state absorption spectrum of anatase TiO$_2$ NPs dispersed in aqueous solution.}
		\label{fig:Absorption_NPs}
	\end{center}
\end{figure}
\newpage

\begin{figure}[h!]
	\centering
	\centering
	\includegraphics[width=\textwidth]{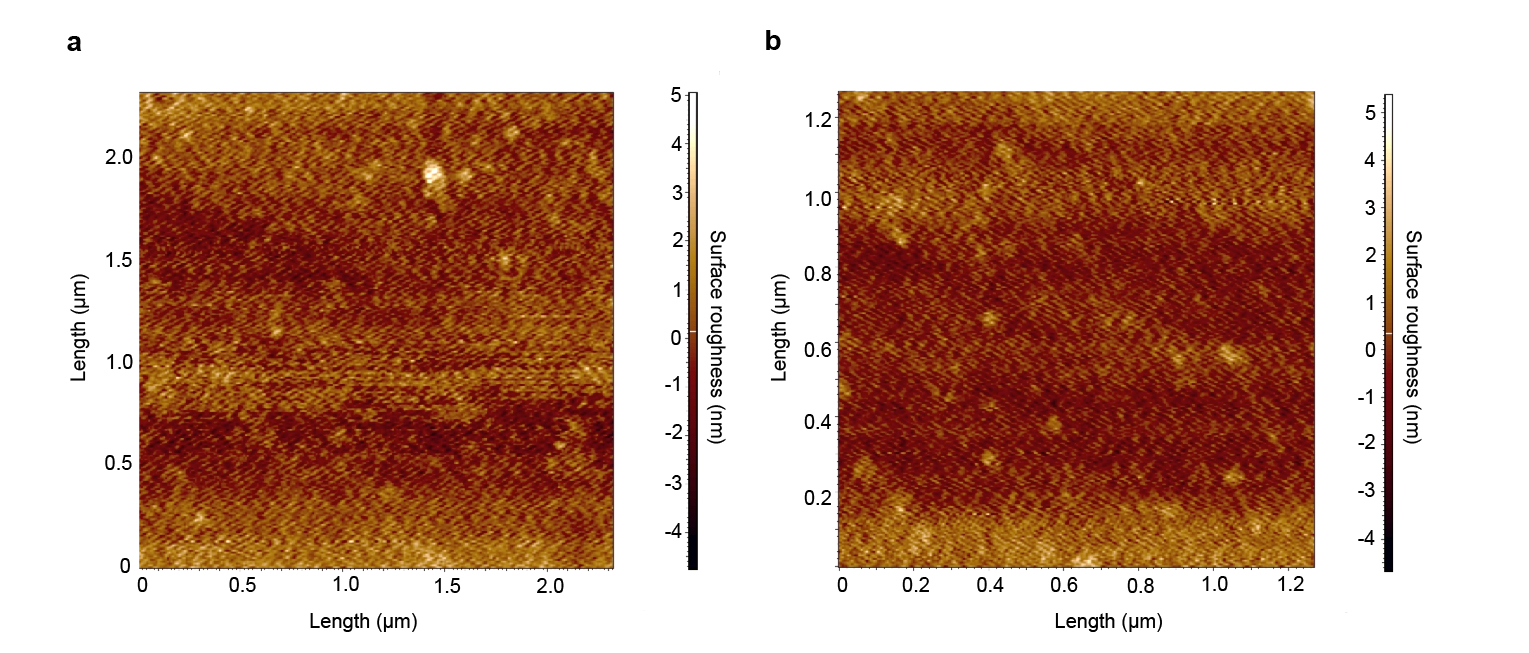}
	\caption{\textbf{Atomic Force Microscopy images of anatase TiO$_2$ single crystals.}\\ \textbf{(a, b)} Roughness characterization of the (010)-oriented polished surface of the reduced anatase TiO$_2$ single crystal used for the SE measurement. The images are taken using Atomic Force Microscopy and the average surface roughness is estimated around 0.9 nm.}
	\label{fig:Fig_AFM}
\end{figure}
\newpage

\begin{figure}[h!]
	\centering
	\centering
	\includegraphics[width=0.6\textwidth]{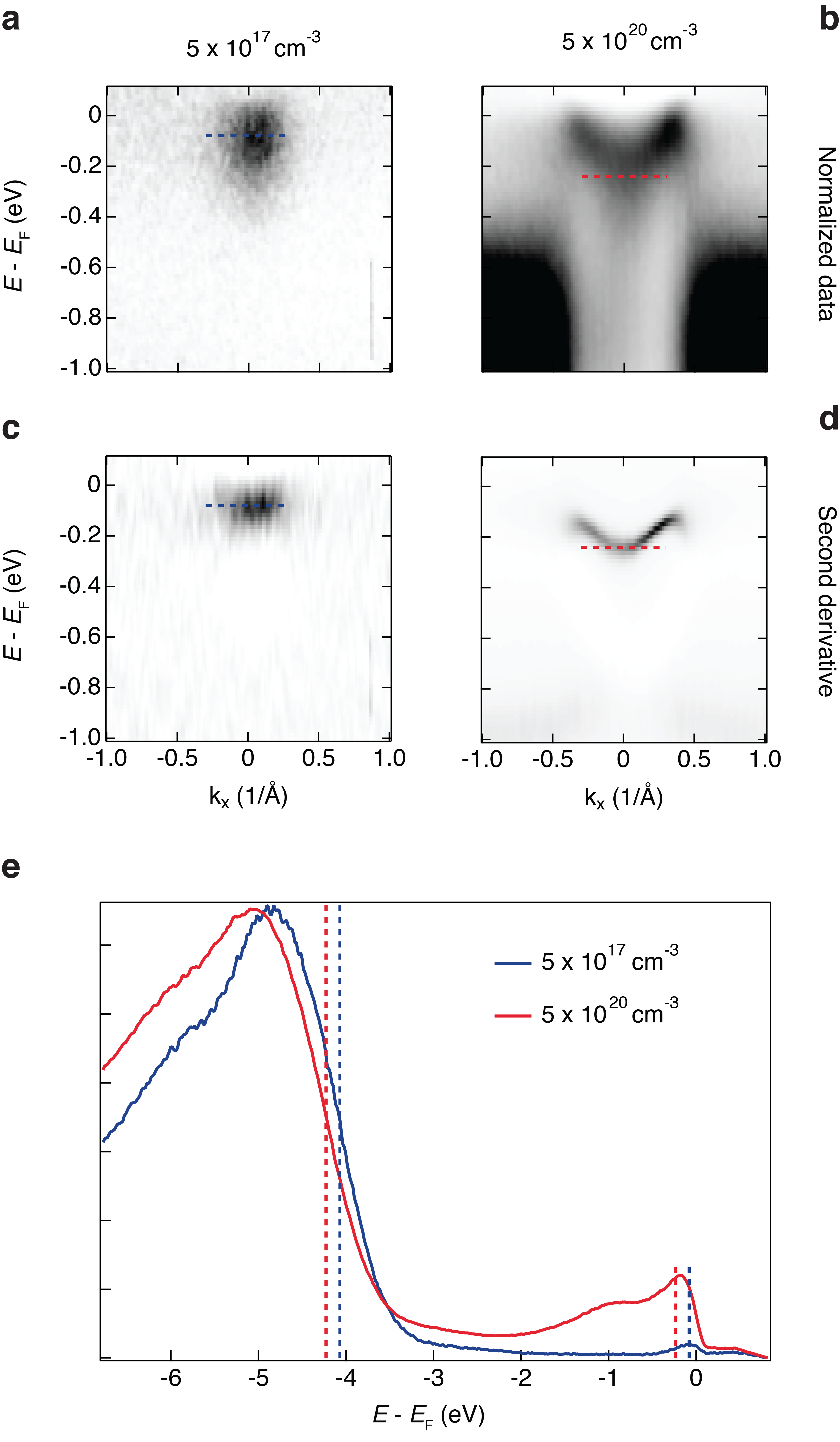}
	\caption{\textbf{Doping dependence of the ARPES data.}\\ \textbf{(a,b)} Energy vs. momentum dispersion of the bottom of the CB for $n$-doped samples with $n$ = 5 $\times$ 10$^{17}$ cm$^{-3}$ and $n$ = 5 $\times$ 10$^{20}$ cm$^{-3}$. \textbf{(c,d)} Second derivative maps of the energy vs. momentum dispersion obtained from panels \textbf{(a,b)}, respectively. \textbf{(e)} Energy distribution curves at the $\Gamma$ point of the BZ for the two considered doping levels. The dashed vertical lines identify the positions of the quasiparticle energies for the VB and CB in the two different samples. The spectra are referenced to $E\mathrm{_F}$.}
	\label{fig:GapARPES}
\end{figure}
\newpage

\begin{figure}[h!]
	\centering
	\centering
	\includegraphics[width=\textwidth]{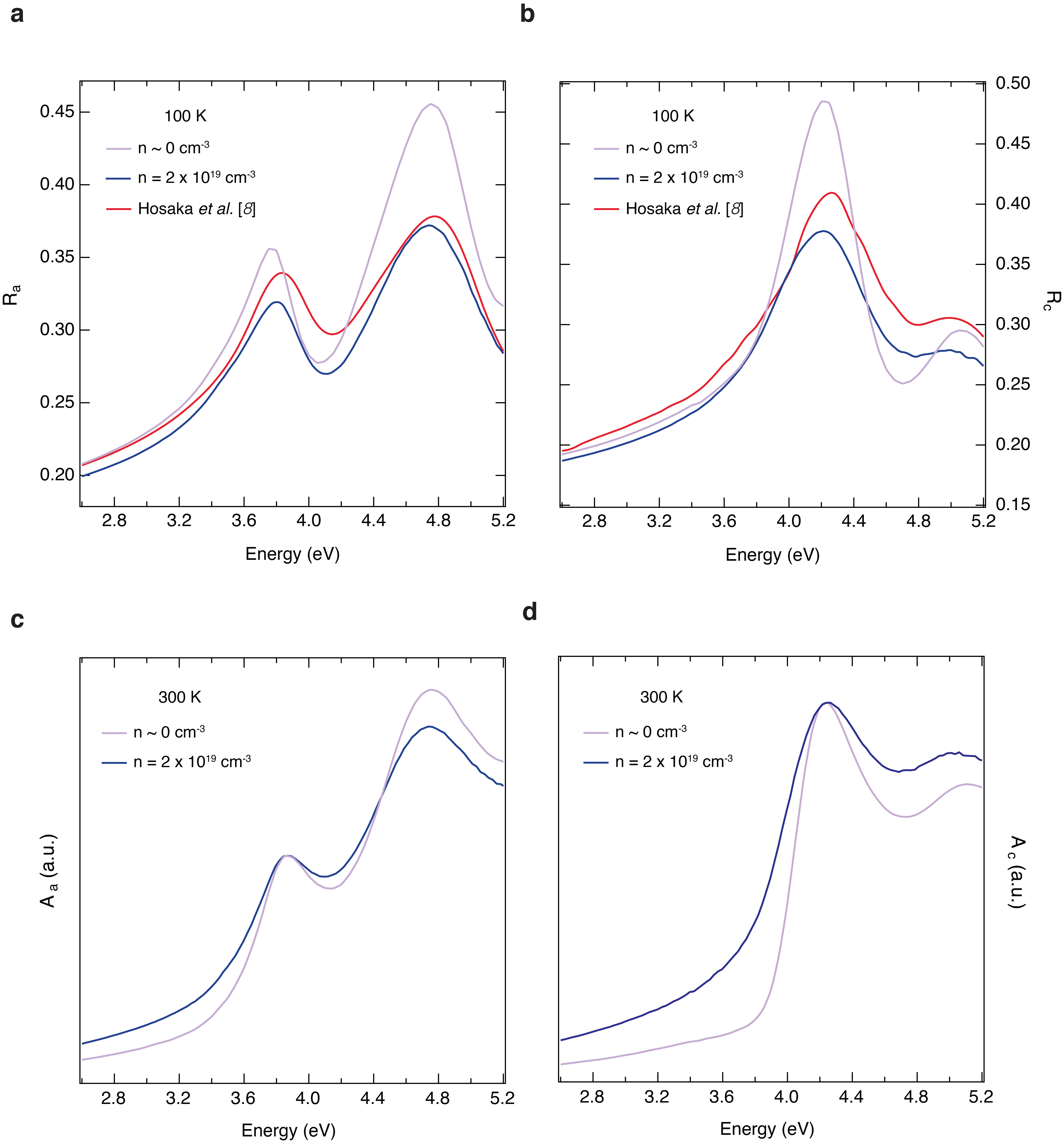}
	\caption{\textbf{Steady-state reflectance and absorption of anatase TiO$_2$ single crystals}.\\\textbf{(a,b)} Steady-state reflectivity spectra of (010)-oriented anatase TiO$_2$ single crystals at 100 K. The electric field is polarized along: \textbf{a} the a-axis; \textbf{b} the c-axis of the crystals. The data derived from our SE measurements are depicted in violet for the pristine ($n$ $\sim$ 0 cm$^{-3}$) crystal and in blue for the $n$-doped ($n$ = 2 $\times$ 10$^{19}$ cm$^{-3}$) crystal, while the reflectivity measured in Ref. 1 is reported in red lines. \textbf{(c,d)}, Normalized steady-state absorption spectra of (010)-oriented anatase TiO$_2$ single crystals at 300 K. The electric field is polarized along: \textbf{c} the a-axis; \textbf{d} the c-axis. The data derived from our SE measurements are depicted in violet for the pristine ($n$ $\sim$ 0 cm$^{-3}$) crystal and in blue for the $n$-doped ($n$ = 2 $\times$ 10$^{19}$ cm$^{-3}$) crystal.}
	\label{fig:FigS2}
\end{figure}
\newpage

\begin{figure}[h!]
	\begin{center}
		\centering
		\includegraphics[width=\textwidth]{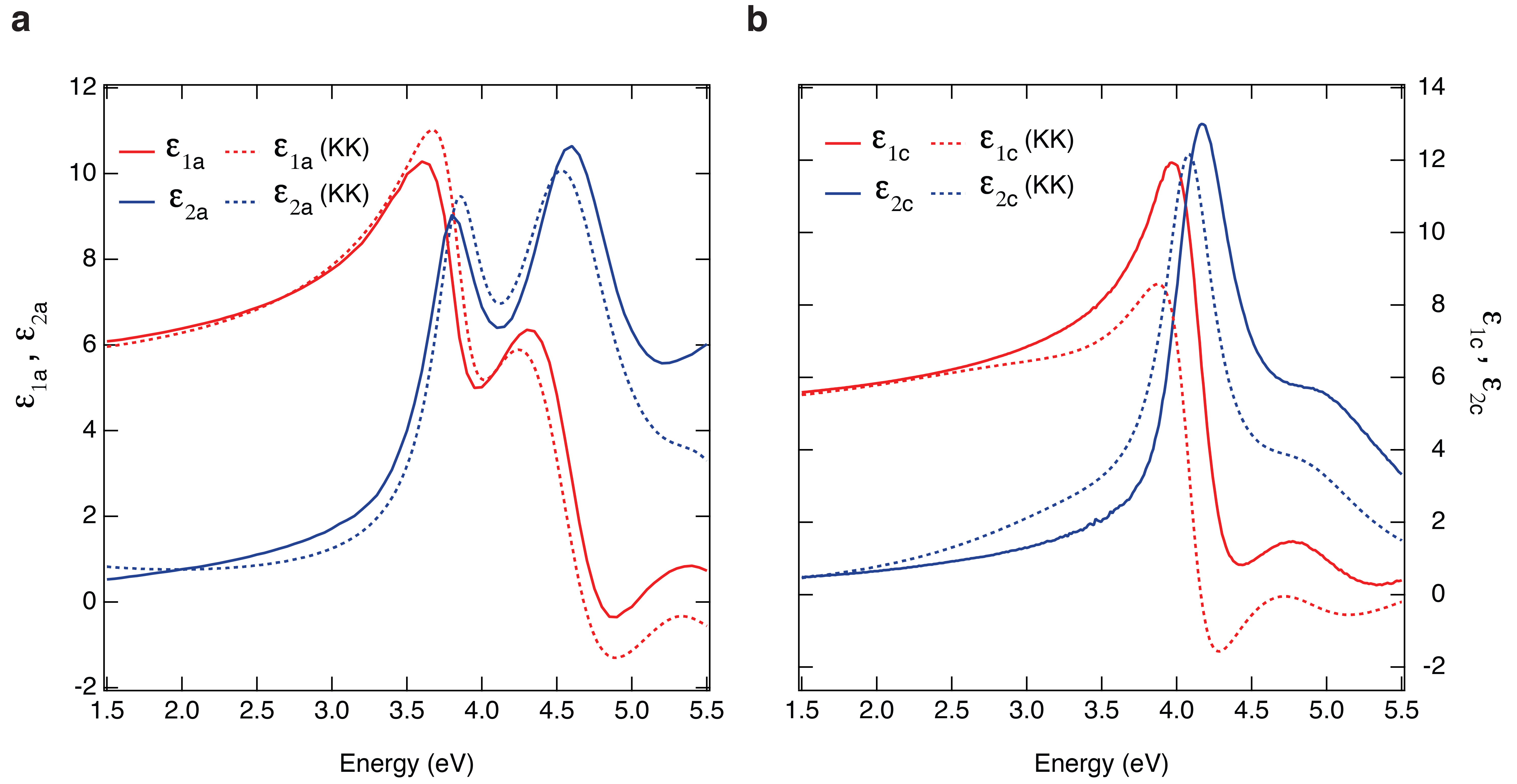}
		\caption{\textbf{Complex dielectric function of anatase TiO$_2$ single crystals at RT.}\\ The electric field is polarized along \textbf{(a)} the a-axis and \textbf{(b)} the c-axis. The real part, $\epsilon_1(\omega)$, is plotted in red, while the imaginary part, $\epsilon_2(\omega)$, in blue. The solid-line curves depict the data directly extracted from SE, while the dashed lines are calculated by a KK analysis of reflectivity.}
		\label{fig:Epsilon_KK}
	\end{center}
\end{figure}
\newpage

\begin{figure}[h!]
	\begin{center}
		\centering
		\includegraphics[width=0.5\textwidth]{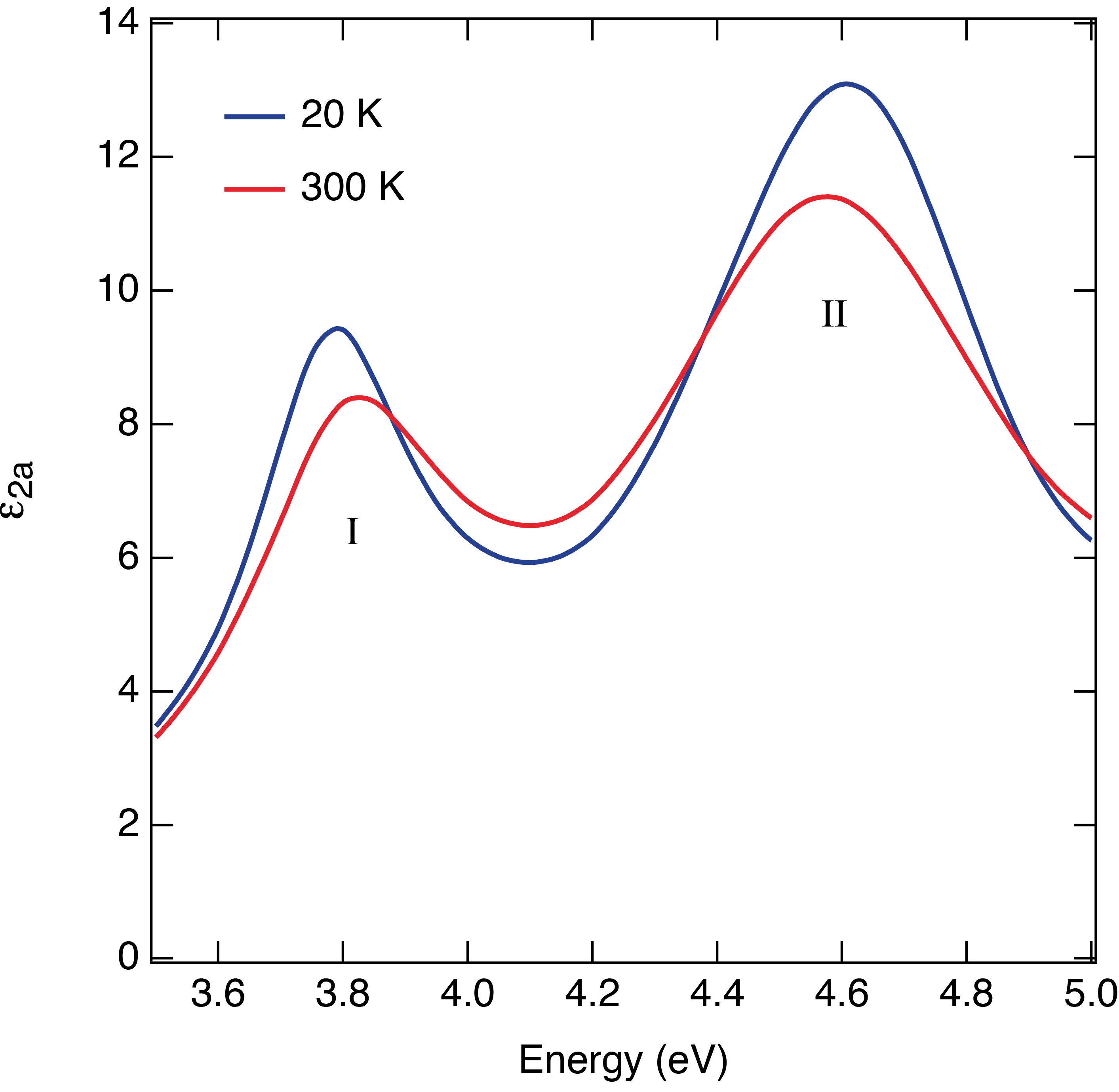}
		\caption{\textbf{Temperature dependence of exciton peak I.}\\ Imaginary part of the dielectric function at 20 K (blue curve) and 300 K (red curve) for \textbf{E} $\perp$ c. The exciton I is observed to undergo an anomalous blueshift of its peak energy, while the charge excitation II displays a conventional redshift.}
		\label{fig:ExcitonTemperature}
	\end{center}
\end{figure}
\newpage

\begin{figure}[h!]
	\begin{center}
		\centering
		\includegraphics[width=0.8\textwidth]{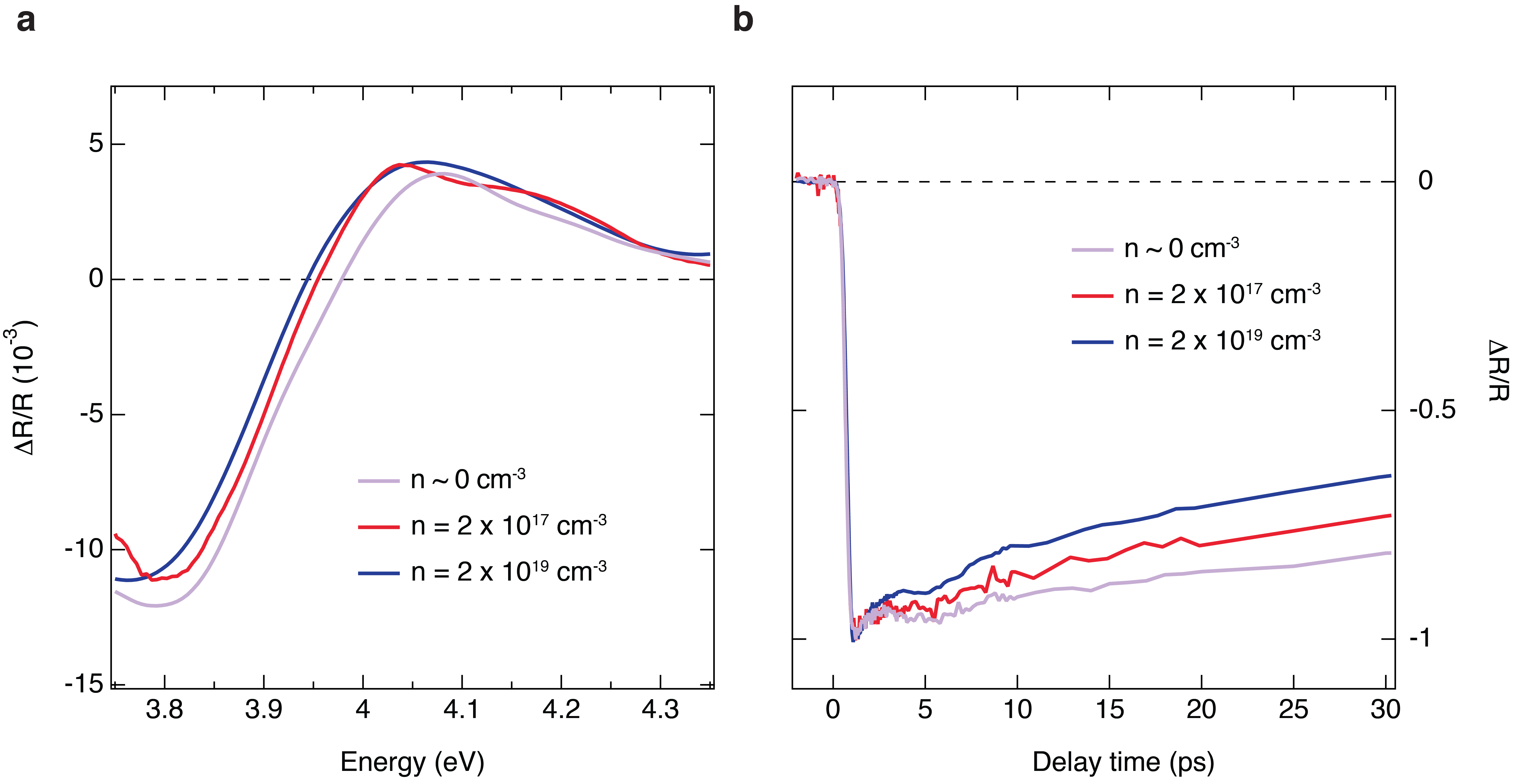}
		\caption{\textbf{Ultrafast broadband UV reflectivity on different classes of anatase TiO$_2$ single crystals at RT.}\\ The doping levels are indicated in the labels. Both pump and probe polarizations lie along the a-axis and the pump photon energy is set at 4.10 eV: \textbf{(a)} Transient spectrum at the fixed time delay of 6 ps. \textbf{(b)} Normalized temporal traces at a fixed probe energy of 3.82 eV.}
		\label{fig:FigS4}
	\end{center}
\end{figure}
\newpage

\begin{figure}[h!]
	\begin{center}
		\centering
		\includegraphics[width=0.8\textwidth]{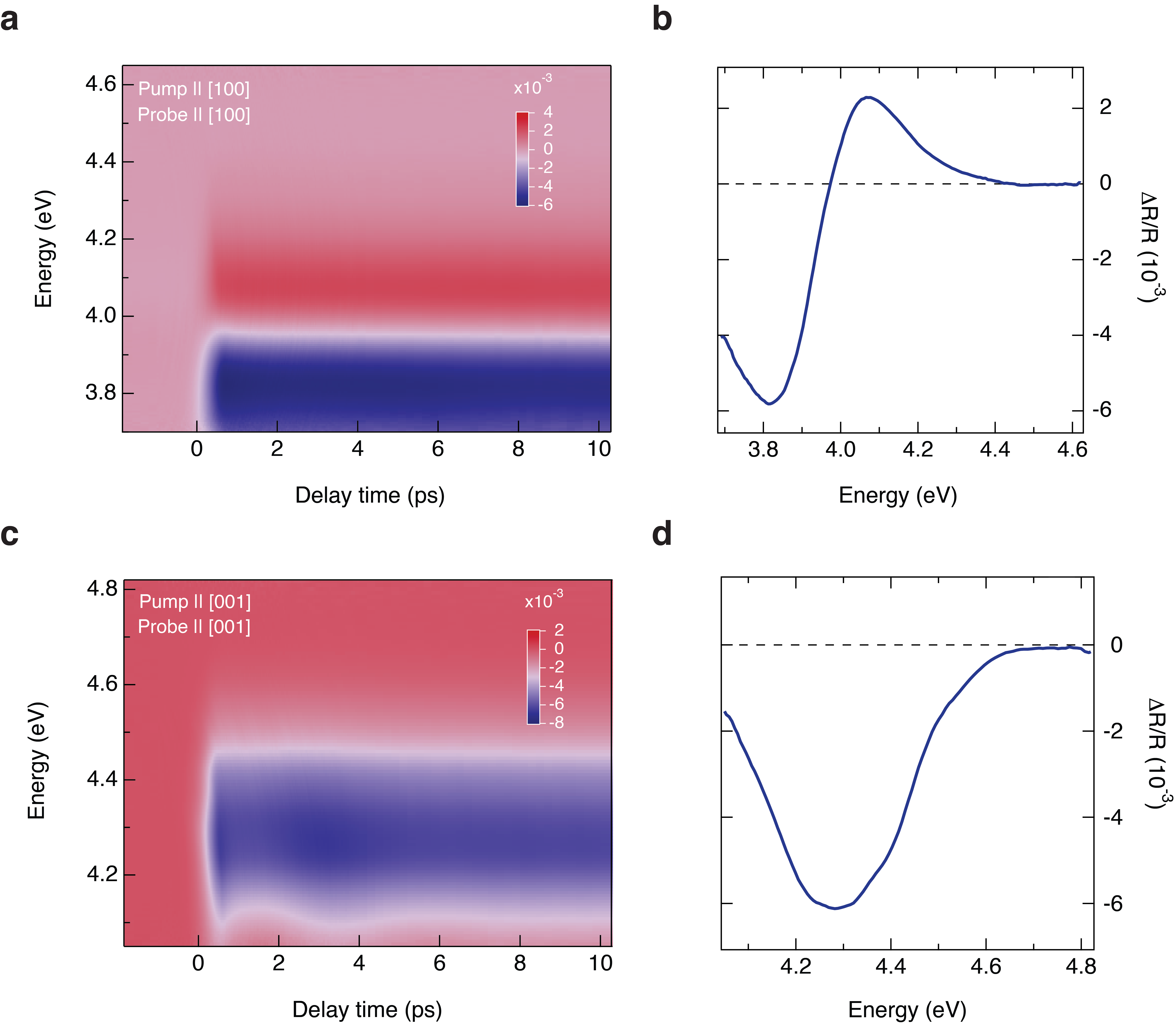}
		\caption{\textbf{Ultrafast anisotropic response of anatase TiO$_2$ single crystals.}\\ Ultrafast broadband UV reflectivity on a (010)-oriented anatase TiO$_2$ single crystal ($n$ = 2 $\times$ 10$^{19}$ cm$^{-3}$) at RT. \textbf{(a,b)} Colour-coded maps of $\Delta R/R$ measured upon photoexcitation at 4.40 eV. \textbf{(c,d)} Transient spectra, obtained from a cut at 1 ps in the experimental conditions reported for \textbf{a} and \textbf{b}, respectively.}
		\label{fig:Composite}
	\end{center}
\end{figure}
\newpage

\begin{figure}[h!]
	\begin{center}
		\centering
		\includegraphics[width=0.5\textwidth]{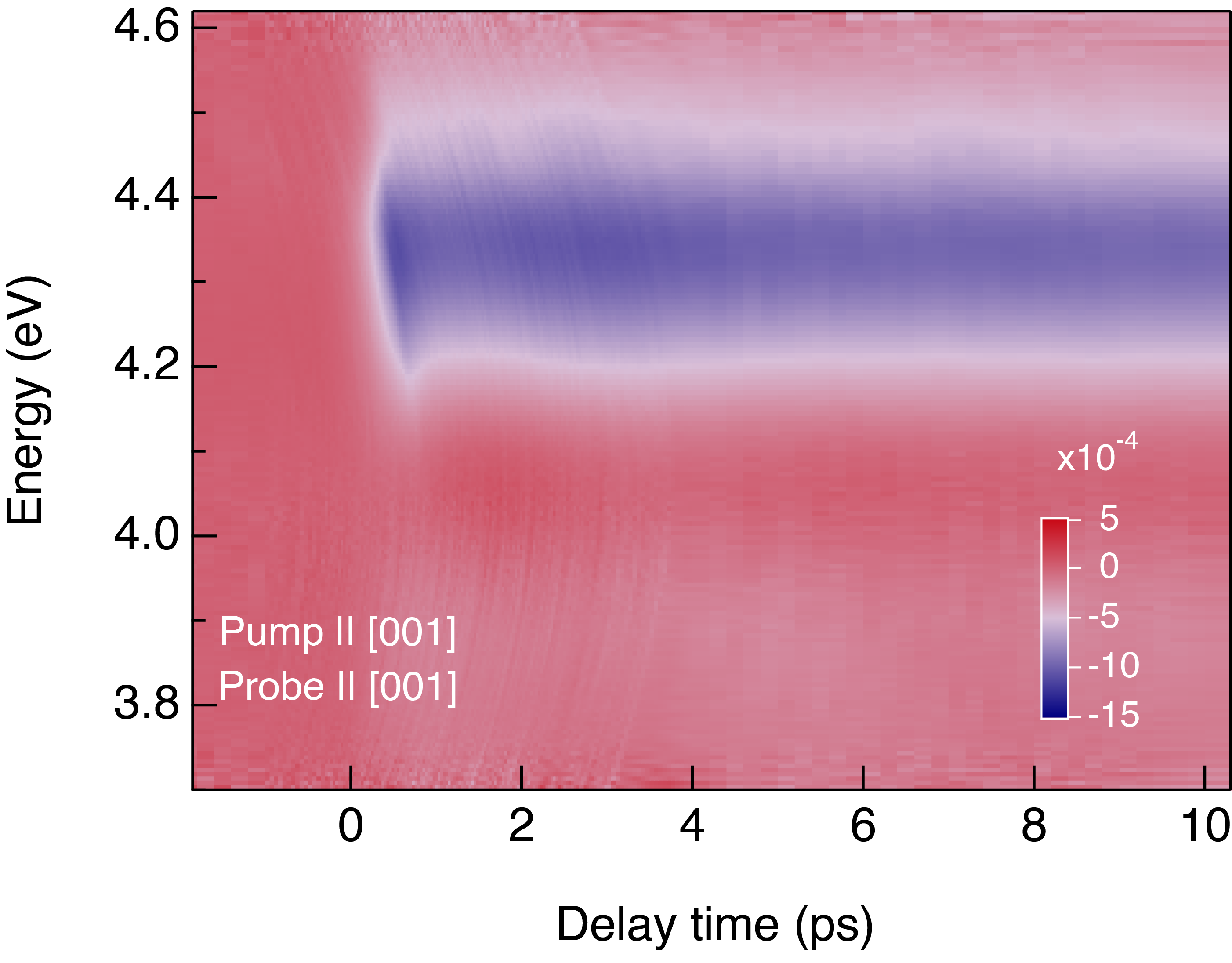}
		\caption{\textbf{Absence of low-energy spectral features in the c-axis ultrafast response.}\\ Colour-coded map of $\Delta R/R$ from a (010)-oriented single crystal ($n$ = 2 $\times$ 10$^{19}$ cm$^{-3}$) measured at RT upon photoexcitation at 4.40 eV and with pump and probe beams polarized along the c-axis. The probe photon energy covers the spectral range 3.70 - 4.60 eV, which demonstrates the absence of emerging features at low energies.}
		\label{fig:TransientReflectivity_caxis}
	\end{center}
\end{figure}
\newpage

\begin{figure}[h!]
	\begin{center}
		\centering
		\includegraphics[width=\textwidth]{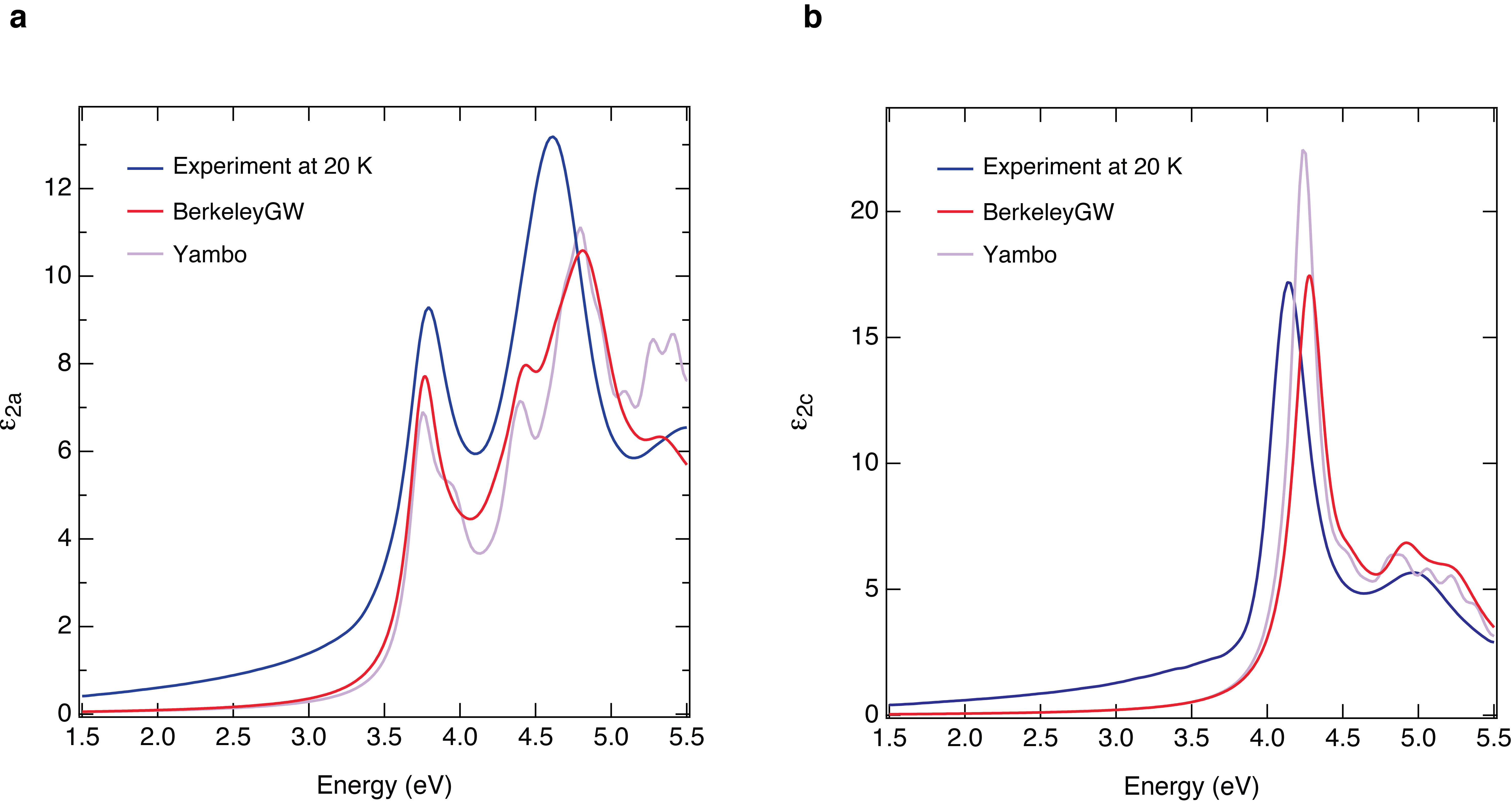}
		\caption{\textbf{Comparison between the experimental SE data and BSE calculations.}\\ Both BerkeleyGW (red curve) and Yambo (violet curve) data are evaluated using the highest convergence parameter values described in the text. Experimental data at 20 K are shown in blue.
			For BerkeleyGW, they correspond to the best converged spectra (both for peaks shape and position). In Yambo, the spectra have been obtained with a less dense $k$-grid, leading the a-axis spectrum to show a spurious shoulder above exciton peak I, as in the previously published works \cite{ref:chiodo}. The fully-converged spectra (red curves) show a single peak in agreement with the experimental data. For light polarized along the a-axis, the agreement between the two calculations is excellent.}
		\label{fig:BSECalculations}
	\end{center}
\end{figure}
\newpage

\begin{figure}[h!]
	\begin{center}
		\centering
		\includegraphics[width=0.5\textwidth]{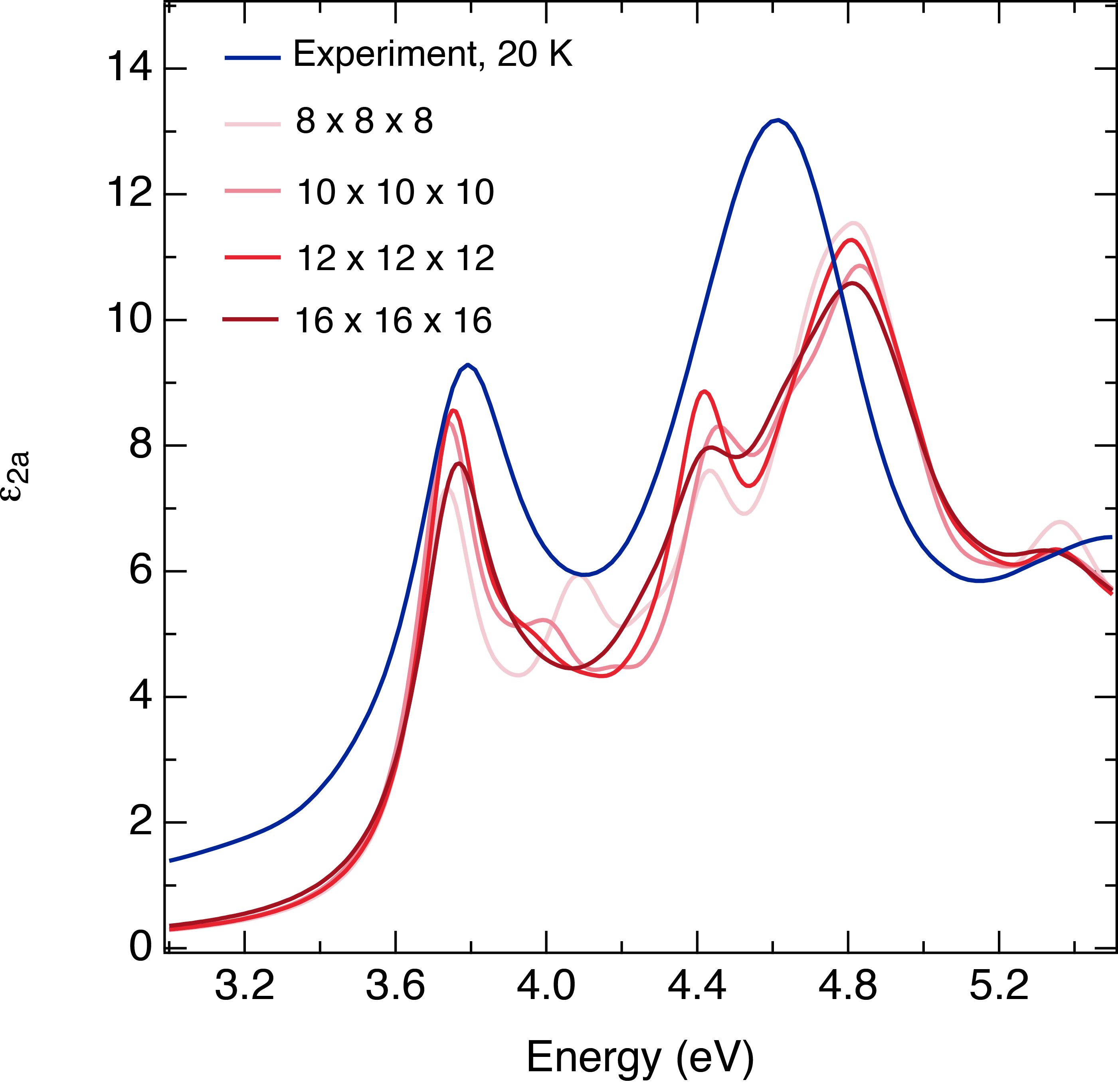}
		\caption{\textbf{Convergence test for the dielectric function.}\\ Convergence test performed with respect to the size of the $k$-points grid using the BerkeleyGW code, in red-colour scale. For comparison, the experimental data at 20 K are also shown (blue curve). To obtain the proper shape of the spectra, it is necessary to use a very large $k$-point grid together with a very large number of bands as described in the text.}
		\label{fig:BGWvsYambo}
	\end{center}
\end{figure}
\newpage

\begin{figure}[h!]
	\begin{center}
		\centering
		\includegraphics[width=0.5\textwidth]{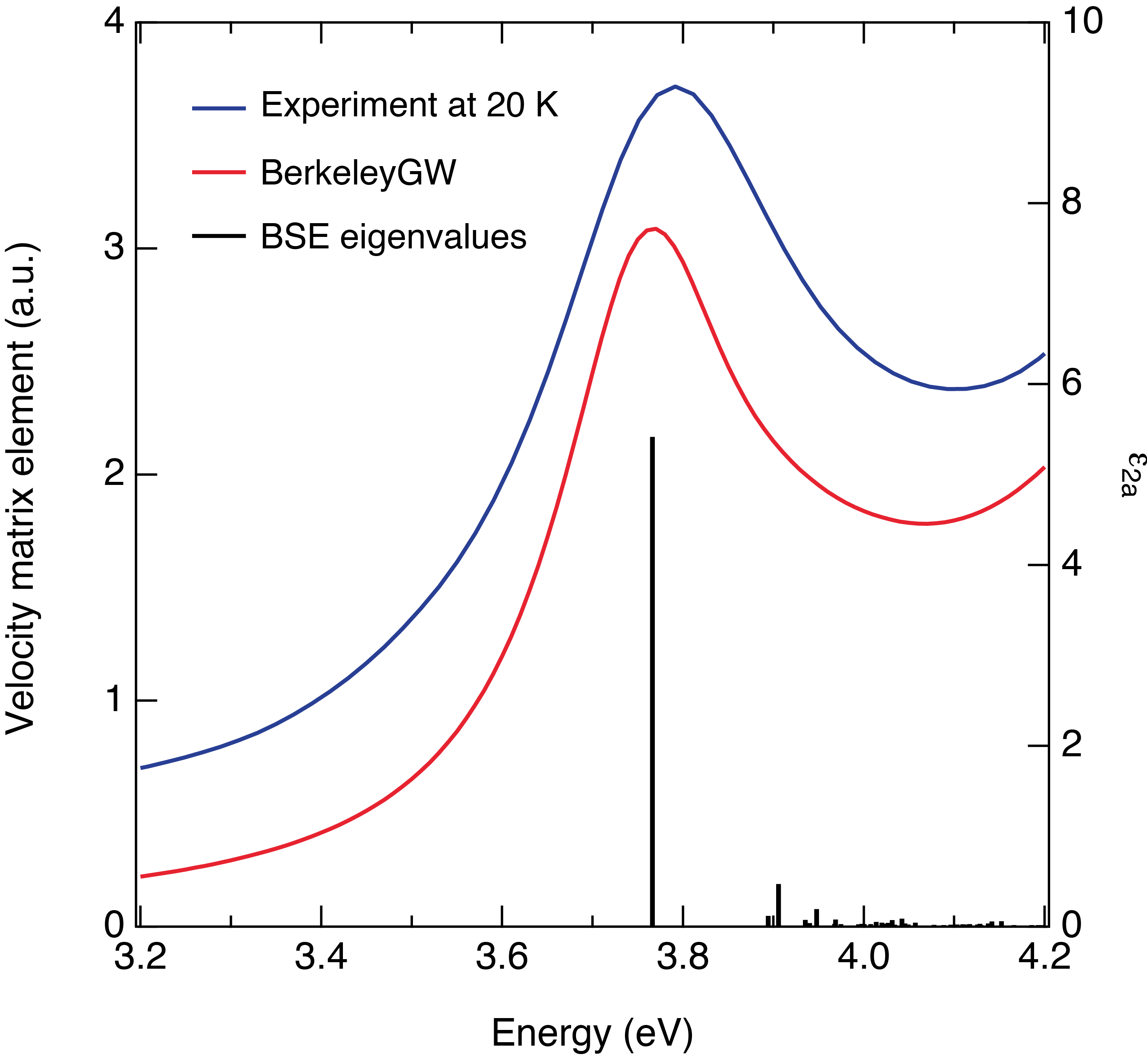}
		\caption{\textbf{Eigenvalue analysis on exciton peak I.}\\ Black bars represent the square of the transition matrix elements of the velocity operator along the a-axis ($|$\textbf{E$\mathrm{_a}$} $\cdot \bra{0}$ \textbf{v} $\ket{\mathrm{S}}|^2$), corresponding to exciton states S contributing to the peak I. This quantity is related to the oscillator strength $f\mathrm{_S}$ by $f\mathrm{_S}$ = (2 $|$ \textbf{E}$\mathrm{_a}$ $\cdot$ $\bra{0}$ \textbf{v} $\ket{S}|^2$)/$E\mathrm{_S}$, where $E\mathrm{_S}$ is the excitation energy corresponding to exciton state S. The experimental $\epsilon_\mathrm{{2a}}$ (blue curve) and the full-converged BerkeleyGW calculations with a phenomenological Lorentzian broadening of 0.12 eV (red curve) are also compared.}
		\label{fig:Eigenvalues}
	\end{center}
\end{figure}
\newpage

\begin{figure}[h!]
	\begin{center}
		\centering
		\includegraphics[width=0.5\textwidth]{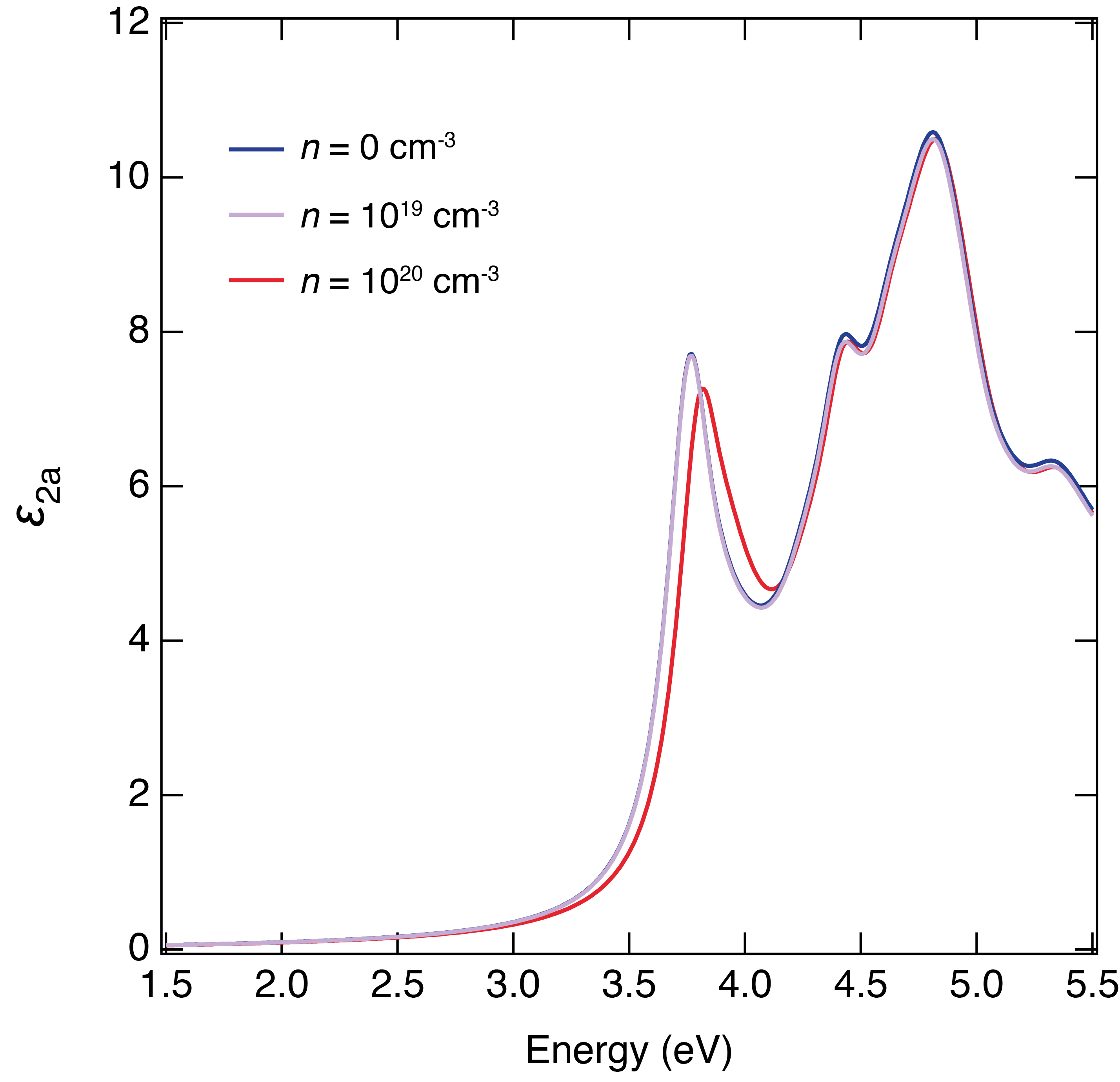}
		\caption{\textbf{Doping dependence of the BSE spectrum.}\\ Comparison between the calculated $\epsilon_\mathrm{{2a}}$ for pristine and $n$-doped anatase TiO$_2$. The optical response of the $n$-doped TiO$_2$ with $n$ = 10$^{19}$ cm$^{-3}$ (violet curve) overlaps almost completely to the pristine case (blue curve), showing that this doping level does not produce a significant effect the peak energy of feature I. Only when $n$ is increased to 10$^{20}$ cm$^{-3}$ (red curve), the peak energy of feature I is found to blueshift of $\sim$ 50 meV}
		\label{fig:DopingDependence}
	\end{center}
\end{figure}
\newpage
\clearpage

\end{document}